\documentclass[usenatbib,letterpaper]{mn2e}
\usepackage[totalwidth=480pt, totalheight=680pt]{geometry}
\usepackage{times}
\usepackage{epsfig}
\usepackage{amsmath}
\usepackage{amssymb}
\usepackage{url}
\usepackage{aas_macros}
\usepackage{color}
\title[Missing dark matter in dwarf galaxies?]{Missing dark matter in dwarf galaxies?}
\author[K. A. Oman et al.]{
\newauthor Kyle A. Oman$^{1,}$\thanks{koman@uvic.ca}, Julio F. Navarro$^{1,2}$, Laura V. Sales$^{3}$, Azadeh Fattahi$^{1}$, \newauthor Carlos S. Frenk$^{4}$, Till Sawala$^{4,5}$, Matthieu Schaller$^{4}$ and Simon D. M. White$^{6}$
\\
$^{1}$ Department of Physics \& Astronomy, University of Victoria, Victoria, BC, V8P 5C2, Canada\\
$^{2}$ Senior CIfAR Fellow\\
$^{3}$ Department of Physics and Astronomy, University of California at Riverside, Riverside, CA, 92521. \\
$^{4}$ Institute for Computational Cosmology, Department of Physics, University of Durham, South Road, Durham DH1 3LE, United Kingdom\\
$^{5}$ Department of Physics, University of Helsinki, Gustaf H\"allstr\"omin katu 2a, FI-00014 Helsinki, Finland\\
$^{6}$ Max Planck Institute for Astrophysics, D-85748 Garching, Germany
}
\date{\today}

\def\eqsim{\mathrel{\raise0.35ex\hbox{$\scriptstyle =$}\kern-0.6em
\lower0.40ex\hbox{{$\scriptstyle \sim$}}}}
\def\gtrsim{\mathrel{\raise0.35ex\hbox{$\scriptstyle >$}\kern-0.6em
\lower0.40ex\hbox{{$\scriptstyle \sim$}}}}
\def\lesssim{\mathrel{\raise0.35ex\hbox{$\scriptstyle <$}\kern-0.6em
\lower0.40ex\hbox{{$\scriptstyle \sim$}}}}

\begin{document}
\label{firstpage}
\maketitle

\begin{abstract} 
  We use cosmological hydrodynamical simulations of the APOSTLE project along with high-quality rotation curve observations to examine the fraction of baryons in $\Lambda$CDM haloes that collect into galaxies. This `galaxy formation efficiency' correlates strongly and with little scatter with halo mass, dropping steadily towards dwarf galaxies. The baryonic mass of a galaxy may thus be used to place a lower limit on total halo mass and, consequently, on its asymptotic maximum circular velocity. A number of observed dwarfs seem to violate this constraint, having baryonic masses up to ten times higher than expected from their rotation speeds, or, alternatively, rotating at only half the speed expected for their mass. Taking the data at face value, either these systems have formed galaxies with extraordinary efficiency -- highly unlikely given their shallow potential wells -- or their dark matter content is much lower than expected from $\Lambda$CDM haloes. This `missing dark matter' is reminiscent of the inner mass deficit of galaxies with slowly-rising rotation curves, but cannot be explained away by star formation-induced `cores' in the dark mass profile, since the anomalous deficit applies to regions larger than the luminous galaxies themselves. We argue that explaining the structure of these galaxies would require either substantial modification of the standard Lambda cold dark matter paradigm or else significant revision to the uncertainties in their inferred mass profiles, which should be much larger than reported. Systematic errors in inclination may provide a simple resolution to what would otherwise be a rather intractable problem for the current paradigm.  \end{abstract}


\section{Introduction}
\label{SecIntro}

The baryon content of the Universe is one of the best known parameters of the present cosmological paradigm, and is well constrained by a variety of independent observations, ranging from the cosmic abundance of the light elements \citep[e.g.,][]{Steigman2007} to the fluctuations in the cosmic microwave background radiation \citep[e.g.,][]{Hu1995,Planck2015}. It is now widely accepted that the Universe has critical density ($\Omega\eqsim 1$) and that matter makes up $\sim 31\%$ of the total matter-energy density ($\Omega_{\rm M}\sim 0.31$), with baryons contributing only a modest fraction \citep[$f_{\rm bar}=\Omega_b/\Omega_{\rm M}\sim 0.17$,][]{Planck2015}. 

Only a fraction of the Universe's baryons are at present locked up within the luminous regions of galaxies: current estimates of this quantity are in the range $\sim 6$-$10\%$ \citep[see, e.g.,][]{Madau2014}. Galaxy formation has thus been a very inefficient process; most of the available baryons have been prevented (or pre-empted) from condensing into galaxies, presumably by cosmic reionization and by the feedback effect of the energetic output of evolving stars and active galactic nuclei. 

A simple quantitative estimate of the resulting galaxy formation efficiency -- which we define hereafter as $f_{\rm eff}=M_{\rm bar}/(f_{\rm bar}\, M_{200})$, i.e., the ratio between the baryonic mass of a galaxy, $M_{\rm bar}$, to the theoretical maximum consistent with the virial\footnote{We define the virial mass, $M_{200}$, as that enclosed by a sphere of mean density $200$ times the critical density of the Universe, $\rho_{\rm crit}=3H^2/8\pi G$. Virial quantities are defined at that radius, and are identified by a `200' subscript.} mass of its host halo \citep{White1993} -- may be obtained by `abundance matching' modelling of the galaxy population. These models indicate that the mean galaxy formation efficiency should be low in haloes of all masses, peaking at $\sim 18\%$ in galaxies of stellar mass of order $3 \times 10^{10}\, {\rm M}_\odot$ and decreasing steeply toward higher and lower masses \citep[see, e.g.,][and references therein]{Behroozi2013}.

The Milky Way sits near the peak of this relation and, at $f_{\rm eff}\sim 0.2$ \citep[for a baryonic mass of order $\sim 5\times 10^{10}$ and a virial mass of $1.5 \times 10^{12}\, {\rm M}_\odot$,][]{Rix2013,Wang2015}, it is considered something of an outlier where galaxy formation has proceeded particularly efficiently. Galaxy formation is expected to be {\it much} less efficient in fainter systems due to the enhanced feedback effects on shallower potential wells \citep{Larson1974,White1978,Efstathiou1992,Bullock2000,Benson2002}, dropping down to essentially zero in haloes with virial masses below $\sim 10^9 \, {\rm M}_\odot$ \citep{Sawala2016a}. 

The steady decline of $f_{\rm eff}$ with decreasing halo mass is now recognized as one of the basic ingredients of galaxy formation models in the Lambda-Cold Dark Matter ($\Lambda$CDM) paradigm, since it serves to reconcile the steeply-rising low-mass end of the CDM halo mass function with the relatively shallow faint-end of the galaxy stellar mass function \citep{White1991}. Assuming that the scatter in the galaxy mass--halo mass relation remains relatively small at low mass, the baryonic mass of a galaxy thus imposes a fairly strict lower limit on the mass of the halo it inhabits and, given the self-similar nature of CDM halo structure \citep[][]{Navarro1997}, on its asymptotic maximum circular velocity. This basic prediction could in principle be readily verified by analysing galaxies where high-quality estimates of their baryonic masses and rotation speeds are available.

A few issues must be considered, however, when attempting such a comparison. Observational estimates of baryonic masses include the contributions of stars and atomic/molecular gas, and are subject to uncertainties in the mass-to-light ratio of the stellar component; in the conversion from neutral hydrogen to total gaseous mass; and in the distance to each individual galaxy (well-studied dwarfs are usually too close for redshift-based distance estimates to be accurate). Another problem is the short radial extent of rotation curves, which in many cases are still rising at the outermost point and, therefore, do not constrain the maximum circular velocity of the system. Finally, observations measure gas velocity fields, which are usually translated into estimates of {\it circular} velocity curves to probe the underlying gravitational potential. This translation includes corrections for inclination, asymmetric drift, non-axisymmetric and random motions, and instrumental limitations which must be carefully taken into account, especially in dwarf galaxies, many of which are notorious for their irregular morphology. 

The theoretical modelling introduces additional uncertainties. A large scatter in galaxy formation efficiency in low-mass haloes might be expected given the sharp decline in $f_{\rm eff}$ required as haloes approach the mass below which galaxies fail to form \citep{Ferrero2012}. In addition, baryons may alter the structure of the dark halo, creating  cores that reduce the central density and depress systematically local estimates of the circular velocity \citep{NavarroEkeFrenk1996,MashchenkoCouchmanWadsley2006,Pontzen2014}.

The observational issues may be addressed by selecting for analysis a galaxy sample with well-calibrated distances, good photometry in multiple passbands, and rotation curves that provide estimates of the circular velocity well beyond the radius that contains the majority of the stars in a galaxy. We therefore focus here on some of the best studied nearby galaxies, including those from (i) the THINGS \citep{Walter2008} and LITTLE THINGS \citep{Hunter2012} surveys; (ii) six dwarfs with exquisite multiwavelength data from \citet{Adams2014}, as well as (iii) those included in the baryonic Tully-Fisher compilation of \citet{McGaugh2012}. The $77$ selected galaxies span nearly four decades in baryonic mass, $10^7<M_{\rm bar}/{\rm M}_\odot<10^{11}$, and roughly a decade in maximum rotation speed, $20<V_{\rm rot}^{\rm max}/$~km~s$^{-1}$~$<200$.

We address the theoretical modelling issues by using results from some of the latest $\Lambda$CDM cosmological hydrodynamical simulations of galaxy formation. We use, in particular, results from the APOSTLE\footnote{APOSTLE stands for `A Project Of Simulating The Local Environment', a suite of $12$ volumes selected from a large cosmological box to match the main properties of the Local Group of Galaxies and its immediate surroundings.}  suite of simulations \citep{Fattahi2015}, which uses the same code developed for the EAGLE project \citep{Schaye2015,Crain2015}. This code, based on {\small P-GADGET3}, a descendent of the {\small GADGET2} code \citep{Springel2005}, has been shown to reproduce the galaxy size and stellar mass functions in a cosmological volume as well as the abundance and properties of dwarf galaxies and satellite systems in the Local Group \citep{Sawala2016b}. These simulations thus provide realistic estimates of the dependence of galaxy formation efficiency on halo mass, as well as its scatter.

Dark matter cores do not develop in dwarfs in the APOSTLE simulations, presumably as a result of choices made when implementing subgrid physics in EAGLE \citep{Schaller2015,Oman2015}. These choices are effective at preventing the artificial fragmentation of gaseous disks, but also limit the magnitude of fluctuations in the gravitational potential that result from the assembly and dispersal of dense star-forming gas clouds. The latter, according to recent work, might lead to the formation of cores in the dark matter \citep{Pontzen2014}. We therefore supplement our analysis with results from the literature where `baryon-induced cores' have been reported \citep{Brook2012,Chan2015,Santos-Santos2016}. 

Like APOSTLE, other simulations have also attempted to reproduce the Local Group environment and kinematics, notably those from the CLUES project \citep{Gottloeber2010} and from the ELVIS project \citep{GarrisonKimmel2014}. We do not include their results here, however, mainly because (i) ELVIS consists of runs that follow solely the dark matter component, and because (ii) the feedback algorithm adopted in CLUES is too weak to prevent excessive star formation in low mass haloes, leading to an unrealistic number of masive dwarfs \citep[see, e.g.,][]{Benitez-Llambay2013}.

We begin by describing the simulated (\S\ref{SecSimGx}) and observed (\S\ref{SecObsGx}) galaxy samples. We then analyse (\S\ref{SecFeff}) the baryon content and galaxy formation efficiency of APOSTLE galaxies and establish their correlations with halo mass/circular velocity. These relations are compared with our observed galaxy sample, an exercise that yields a number of outliers for which there are no counterparts in the simulations. Particularly interesting are outliers inferred to have exceptionally high galaxy formation efficiency, or, alternatively, to rotate far too slowly for their baryonic mass, presumably because they are anomalously deficient in dark matter. Neither possibility finds a natural explanation in current simulations of dwarf galaxy formation. We examine in \S\ref{SecRotCHM} the possibility that this issue is related to the question of cores inferred in the inner rotation curves of some dwarf galaxies, and whether errors in the rotation curve modelling could be the source of the observed anomalies. We conclude in \S\ref{SecConc} with a brief summary and discussion of the implications of these puzzling systems for our understanding of dwarf galaxy formation in a $\Lambda$CDM universe. 

\section{The APOSTLE project}
\label{SecSimGx}

\subsection{The numerical simulations}
\label{SecNumSims}

We select galaxies from the APOSTLE suite of zoom-in hydrodynamical simulations. These follow a total of $12$ volumes specifically selected from a cosmological dark matter-only simulation to contain two haloes with approximately the masses and dynamics of the Milky Way and M~31, and no other nearby large structures \citep[for details, see][]{Fattahi2015,Sawala2016b}. 

APOSTLE uses the same code and physics as the `Ref' EAGLE simulations described by \citet{Schaye2015}. EAGLE uses the pressure-entropy formulation of smoothed particle hydrodynamics \citep{Hopkins2013} and the {\small ANARCHY} collection of numerical methods (Dalla Vecchia et al., in preparation; for a brief description see \citealt{Schaye2015}). It includes subgrid models for radiative cooling \citep{Wiersma2009a}, star formation \citep{Schaye2004,Schaye2008}, stellar and chemical enrichment \citep{Wiersma2009b}, energetic stellar feedback \citep{DallaVecchia2012}, and cosmic reionization \citep{Haardt2001,Wiersma2009b}, and is calibrated to reproduce the galaxy stellar mass function and size distribution for galaxies of $M_*>10^8\,{\rm M}_\odot$ \citep{Crain2015}.

The APOSTLE volumes are simulated at three different resolution levels which we denote AP-L1, AP-L2 and AP-L3 in order of decreasing resolution. Each resolution level is separated by a factor of $\sim 10$ in particle mass and a factor of $\sim 2$ in force resolution. All $12$ volumes have been simulated at AP-L2 and AP-L3 resolution levels, but only volumes $1$ and $4$ have been simulated at AP-L1 resolution. APOSTLE assumes \emph{WMAP7} \citep{WMAP7} cosmological parameters: $\Omega_m = 0.2727$, $\Omega_\Lambda = 0.728$, $\Omega_b = 0.04557$, $h = 0.702$ and $\sigma_8 = 0.807$. Table~\ref{APparams} summarizes the particle masses and softening lengths of each resolution level.

\begin{table}
  \caption{Summary of the key parameters of the APOSTLE simulations used in this work. Particle masses vary by up to a factor of $2$ between volumes at a fixed resolution `level'; the median values below are indicative only \citep[see][for full details]{Fattahi2015}. Details of the \emph{WMAP7} cosmological parameters used in the simulations are available in \citet{WMAP7}.\label{APparams}}
  \begin{tabular}{llll}
    \hline
    & \multicolumn{2}{l}{Particle masses (${\rm M}_\odot$)} & Max softening \\
    Simulation & DM & Gas & length (pc)\\
    \hline
    AP-L3 & $7.3\times10^6$ & $1.5\times10^6$ & $711$ \\
    AP-L2 & $5.8\times10^5$ & $1.2\times10^5$ & $307$ \\
    AP-L1 & $3.6\times10^4$ & $7.4\times10^3$ & $134$ \\
    \hline
  \end{tabular}
\end{table}

\subsection{The simulated galaxy sample}

Galaxies are identified in APOSTLE using the {\small SUBFIND} algorithm \citep{Springel2001,Dolag2009}. Particles are first grouped into friends-of-friends (FoF) haloes by linking together dark matter particles separated by less than 0.2 times the mean inter-particle spacing \citep{Davis1985}; gas and star particles are assigned to the same FoF halo as their nearest dark matter particle within the linking length. Substructures are then separated along saddle points in the density distribution; in this step, dark matter, gas and star particles are treated as a single distribution of mass. Finally, particles that are not gravitationally bound to the substructures are removed. 

We retain for analysis the main (central) galaxy of each separate FoF halo; this excludes by construction satellites of more massive systems and are best identified with `isolated' field galaxies. For each of these galaxies we measure the virial mass of its surrounding halo, $M_{200}$, as well as its baryonic mass, $M_{\rm bar}$, which we identify with the total mass of baryons within the galactic radius, $r_{\rm gal}=0.15\, r_{200}$. This definition includes the great majority of stars and cold gas within the halo virial radius.

We shall consider two characteristic circular velocities for each galaxy in our analysis: (i) the maximum circular velocity, $V_{\rm max}$, measured within the virial radius; and (ii) the velocity at the outskirts of the luminous galaxy, which we identify with the circular velocity at twice the stellar half-mass radius, $V_{\rm circ}(2\, r_h^{st})$. For simplicity, we estimate all circular velocities using the total enclosed mass, assuming spherical symmetry; i.e., $V_{\rm circ}^2(r)=GM(<r)/r$.

We use the three APOSTLE resolution levels to determine which simulated galaxies are sufficiently resolved to measure baryonic masses and circular velocities. We retain AP-L1 galaxies with $V_{\rm circ}^{\rm max}>26$~km~s$^{-1}$, AP-L2 galaxies with $V_{\rm circ}^{\rm max}>56$~km~s$^{-1}$ and AP-L3 galaxies with $V_{\rm circ}^{\rm max}>120$~km~s$^{-1}$ in our sample. These cuts correspond to virial masses of $\gtrsim 3\times10^{9}$, $3\times10^{10}$ and $3\times10^{11}\,{\rm M}_\odot$, respectively, or a particle count $\gtrsim 5\times10^{4}$. All circular velocities used in our analysis are well resolved according to the criterion of \citet{Power2003}.

\section{The observed galaxy sample}
\label{SecObsGx}

Our observed galaxy sample has been drawn from several heterogeneous sources, placing an emphasis on galaxies with good estimates of their baryonic masses and high-quality rotation curves derived from 2D velocity fields. This is a subset of the compilation of rotation curves presented in \citet{Oman2015}, and contains galaxies taken from the sources listed below. We take baryonic masses directly from the listed sources\footnote{We have adopted $M_{\rm gas}/M_{\rm HI}=1.4$ to account for the gas mass in Helium and heavy elements.}, and adopt their published circular velocity estimates, which are based on folded rotation curves corrected for inclination, asymmetric drift, and instrumental effects. No further processing of these data has been attempted. The properties of galaxies in our compilation that have rotation curves extending to at least twice their stellar half mass radius (see \S\ref{SubsecCores}) are summarized in Table~\ref{ObsTable}. Below, we briefly discuss each of these datasets.

\subsection{THINGS and LITTLE THINGS}
\label{SecTHINGS}

Rotation curves for 44 galaxies in the THINGS and LITTLE THINGS surveys were published by \citet{deBlok2008}, \citet{Oh2011} and \citet{Oh2015}. These galaxies span a wide range of masses, with maximum circular velocities between $\sim 20$ and $\sim 400$~km~s$^{-1}$. The surveys obtained ${\rm H}\,\textsc{i}$ data cubes using the NRAO Very Large Array with angular resolutions of $12$ (THINGS) and $6$ (LITTLE THINGS) arcsec, making them some of the most finely spatially resolved ${\rm H}\,\textsc{i}$ rotation curves available. The rotation curves were constructed from the velocity fields using a tilted-ring model \citep{Rogstad1974,Kamphuis2015}, corrected for inclination, and asymmetric drift when necessary. A few galaxies are analysed in multiple publications; in these cases we use only the most recent analysis. ${\rm H}\,\textsc{i}$ masses are derived from the THINGS and LITTLE THINGS data by \citet{Walter2008} and \citet{Oh2015}, respectively. 

Stellar masses are estimated by fitting stellar population spectral energy density models to Spitzer IRAC $3.6\,\mu {\rm m}$ observations \citep{Hunter2006}. We use the disk scale lengths reported in \citet{Hunter2012} to estimate $r_h^{st}$ for LITTLE THINGS galaxies -- for an exponential profile the half mass radius is related to the scale length, $r_d$, as $r_h^{st}\approx1.68\, r_d$. For the THINGS sample, no scale lengths are reported, but the contribution of stars to the circular velocity is shown as a function of radius. We therefore assume an exponential disk profile and estimate a scale length from the position of the peak of the contribution of the stellar component of each galaxy \citep[][\S2.6.1b]{BT2008}.

\subsection{\citet{Adams2014}}
\label{SecAdams}

\citet{Adams2014} present a sample of 7 rotation curves of galaxies with maximum circular velocities of $\sim 100$~km~s$^{-1}$. The velocity fields were measured with the VIRUS-W integral field spectrograph on the 2.7-m Harlan J. Smith Telescope at McDonald Observatory with an angular resolution of $3.1$ arcsec. The authors analyse separately absorption lines, tracing the stellar velocity field, and ${\rm H}$~$\beta$, ${\rm O}\,\textsc{iii}\;4959\,{\textrm \AA}$ and ${\rm O}\,\textsc{iii}\;5007\,{\textrm \AA}$ emission, tracing the gas velocity field. Using a tilted-ring model, two independent rotation curves, one for each velocity field, were constructed for each galaxy. In most cases the two curves are in good agreement. We use the gas emission based curves in our analysis, and note that using the stellar absorption based curves would not change anything substantial in our analysis. We use the disk scale lengths reported by the authors to estimate $r_h^{st}$, and the ${\rm H}\,\textsc{i}$ masses they quote from \citet{Paturel2003}. We use the stellar masses they derive by modelling the gas rotation curves, which are better constrained than those derived by modelling the stellar rotation curves (see, e.g., their fig.~13).

\subsection{\citet{McGaugh2012}}
\label{SecMcGaugh}

We use the compilation of 47 galaxies of \citet{McGaugh2012} to supplement our own compilation. It provides self-consistent estimates of the height of the flat portion of the rotation curve (which we consider equivalent to $V_{\rm rot}^{\rm max}$ in our notation), stellar masses, and gas masses. The gas masses assume $M_{\rm gas}/M_{\rm HI}=1.33$; we increase the gas masses by $\sim 5\%$ for consistency with the rest of our compilation. We remove 7 galaxies already included in our compilation from the THINGS survey and one duplicate entry (UGC 4115 a.k.a. LSB D631-7). The majority of the remaining galaxies do not have high quality rotation curve measurements that are readily available, so we only use these data in our baryonic Tully-Fisher and $f_{\rm eff}$ analysis below.

\begin{table*}
  \caption{Summary of properties for galaxies with rotation curves extending to at least $2r_h^{st}$, ordered by $V_{\rm rot}(2r_h^{st})$, i.e. left-to-right in Fig.~\ref{FigMbarVhalf}. Columns: {\bf(1)} galaxy name used by reference in (2); {\bf(2)} rotation curve source; {\bf(3)} distance as given by reference in (2); {\bf(4)} inclination as given by reference in (2); {\bf(5)} stellar half mass radius estimated as described in \S\ref{SecObsGx}; {\bf(6)} maximum measured rotation velocity; {\bf(7)} measured rotation velocity at twice the stellar half mass radius; {\bf(8)} stellar mass as given by reference in (2); {\bf(9)} baryonic mass assuming stellar mass in (7) and $M_{\rm gas}/M_{\rm HI}=1.4$; {\bf(10)} galaxy formation efficiency as shown in Fig.~\ref{FigFeffVmax}. \label{ObsTable}}
  \begin{tabular}{llrrrrrllr}
    \hline
    \multicolumn{1}{c}{Galaxy} & \multicolumn{1}{c}{Ref.} & \multicolumn{1}{c}{$D$} & \multicolumn{1}{c}{$i$} & \multicolumn{1}{c}{$r_h^{st}$} & \multicolumn{1}{c}{$V_{\rm rot}^{\rm max}$} & \multicolumn{1}{c}{$V_{\rm rot}(2r_h^{st})$} & \multicolumn{1}{c}{$M_{*}$} & \multicolumn{1}{c}{$M_{\rm bar}$} & \multicolumn{1}{c}{$f_{\rm eff}$} \\
    & & \multicolumn{1}{c}{$[{\rm Mpc}]$} & \multicolumn{1}{c}{$[^\circ]$} & \multicolumn{1}{c}{$[{\rm kpc}]$} & \multicolumn{1}{c}{$[{\rm km}\,{\rm s}^{-1}]$} & \multicolumn{1}{c}{$[{\rm km}\,{\rm s}^{-1}]$} & \multicolumn{1}{c}{$[{\rm M}_\odot]$} & \multicolumn{1}{c}{$[{\rm M}_\odot]$} &  \\
    \hline
    IC 1613 & \citet{Oh2015} & $0.7$ & $48$ &$0.97$ &$21.1$ & $19.3$ & $2.88\times10^{7}$ &$8.77\times10^{7}$ & $36.4\%$ \\
NGC 1569 & \citet{Oh2015} & $3.4$ & $69$ &$0.64$ &$39.3$ & $23.0$ & $3.63\times10^{8}$ &$5.67\times10^{8}$ & $34.2\%$ \\
CVnIdwA & \citet{Oh2015} & $3.6$ & $66$ &$0.96$ &$26.4$ & $24.1$ & $4.90\times10^{6}$ &$3.37\times10^{7}$ & $7.0\%$ \\
DDO 43 & \citet{Oh2015} & $7.8$ & $41$ &$0.69$ &$38.3$ & $25.6$ & \multicolumn{1}{c}{---} &$2.34\times10^{8}$ & $15.3\%$ \\
UGC 8508 & \citet{Oh2015} & $2.6$ & $82$ &$0.45$ &$46.1$ & $26.0$ & $7.76\times10^{6}$ &$1.98\times10^{7}$ & $0.7\%$ \\
DDO 50 & \citet{Oh2015} & $3.4$ & $50$ &$1.85$ &$38.8$ & $29.0$ & $1.07\times10^{8}$ &$1.43\times10^{9}$ & $88.9\%$ \\
Haro 29 & \citet{Oh2015} & $5.9$ & $61$ &$0.49$ &$43.5$ & $33.1$ & $1.45\times10^{7}$ &$1.08\times10^{8}$ & $4.7\%$ \\
DDO 70 & \citet{Oh2015} & $1.3$ & $50$ &$0.81$ &$43.9$ & $33.7$ & $1.95\times10^{7}$ &$5.75\times10^{7}$ & $2.4\%$ \\
LSB F564-V3 & \citet{Oh2015} & $8.7$ & $56$ &$0.89$ &$39.2$ & $33.8$ & \multicolumn{1}{c}{---} &$4.37\times10^{7}$ & $2.6\%$ \\
WLM & \citet{Oh2015} & $1.0$ & $74$ &$0.96$ &$38.5$ & $34.3$ & $1.62\times10^{7}$ &$9.57\times10^{7}$ & $6.1\%$ \\
DDO 154 & \citet{Oh2015} & $3.7$ & $68$ &$0.99$ &$51.1$ & $35.9$ & $8.32\times10^{6}$ &$3.63\times10^{8}$ & $9.6\%$ \\
DDO 126 & \citet{Oh2015} & $4.9$ & $65$ &$1.46$ &$38.7$ & $38.7$ & $1.62\times10^{7}$ &$1.78\times10^{8}$ & $11.2\%$ \\
Haro 36 & \citet{Oh2015} & $9.3$ & $70$ &$1.16$ &$58.2$ & $39.5$ & \multicolumn{1}{c}{---} &$1.12\times10^{8}$ & $2.0\%$ \\
DDO 87 & \citet{Oh2015} & $7.7$ & $56$ &$2.20$ &$56.6$ & $44.4$ & $3.24\times10^{7}$ &$3.21\times10^{8}$ & $6.2\%$ \\
NGC 2366 & \citet{Oh2015} & $3.4$ & $63$ &$2.28$ &$59.8$ & $55.5$ & $6.92\times10^{7}$ &$1.14\times10^{9}$ & $18.6\%$ \\
DDO 47 & \citet{Oh2015} & $5.2$ & $46$ &$2.30$ &$64.7$ & $60.1$ & \multicolumn{1}{c}{---} &$4.68\times10^{8}$ & $6.0\%$ \\
DDO 52 & \citet{Oh2015} & $10.3$ & $43$ &$2.18$ &$61.7$ & $60.5$ & $5.37\times10^{7}$ &$3.85\times10^{8}$ & $5.7\%$ \\
DDO 168 & \citet{Oh2015} & $4.3$ & $46$ &$1.38$ &$61.9$ & $60.5$ & $5.89\times10^{7}$ &$3.16\times10^{8}$ & $4.6\%$ \\
NGC 5204 & \citet{Adams2014} & $3.2$ & $47$ &$0.79$ &$89.4$ & $76.2$ & $2.51\times10^{8}$ &$5.33\times10^{8}$ & $2.5\%$ \\
IC 2574 & \citet{Oh2011} & $4.0$ & $55$ &$5.23$ &$80.0$ & $78.2$ & $1.02\times10^{9}$ &$2.84\times10^{9}$ & $18.7\%$ \\
NGC 2552 & \citet{Adams2014} & $11.4$ & $53$ &$3.23$ &$96.1$ & $95.7$ & $1.26\times10^{9}$ &$2.17\times10^{9}$ & $8.1\%$ \\
UGC 11707 & \citet{Adams2014} & $15.0$ & $73$ &$3.69$ &$103.7$ & $96.7$ & $1.20\times10^{9}$ &$3.20\times10^{9}$ & $9.3\%$ \\
NGC 7793 & \citet{deBlok2008} & $3.9$ & $50$ &$2.65$ &$117.9$ & $114.1$ & $2.75\times10^{9}$ &$3.98\times10^{9}$ & $7.8\%$ \\
NGC 2403 & \citet{deBlok2008} & $3.2$ & $63$ &$2.40$ &$143.9$ & $122.7$ & $5.13\times10^{9}$ &$8.76\times10^{9}$ & $9.2\%$ \\
NGC 3621 & \citet{deBlok2008} & $6.6$ & $65$ &$3.83$ &$159.2$ & $139.6$ & $1.58\times10^{10}$ &$2.58\times10^{10}$ & $19.9\%$ \\
NGC 4736 & \citet{deBlok2008} & $4.7$ & $41$ &$2.62$ &$198.3$ & $153.1$ & $2.00\times10^{10}$ &$2.05\times10^{10}$ & $8.0\%$ \\
NGC 3198 & \citet{deBlok2008} & $13.8$ & $72$ &$5.60$ &$158.7$ & $153.4$ & $2.51\times10^{10}$ &$3.92\times10^{10}$ & $30.5\%$ \\
NGC 6946 & \citet{deBlok2008} & $5.9$ & $33$ &$5.34$ &$224.3$ & $195.3$ & $6.31\times10^{10}$ &$6.89\times10^{10}$ & $18.2\%$ \\

    \hline
  \end{tabular}
\end{table*}

\section{Galaxy baryonic mass and dark halo mass}
\label{SecFeff}

\subsection{The baryonic Tully-Fisher relation}
\label{SubsecBTFR}

\begin{figure}
  {\leavevmode \includegraphics[width=\columnwidth]{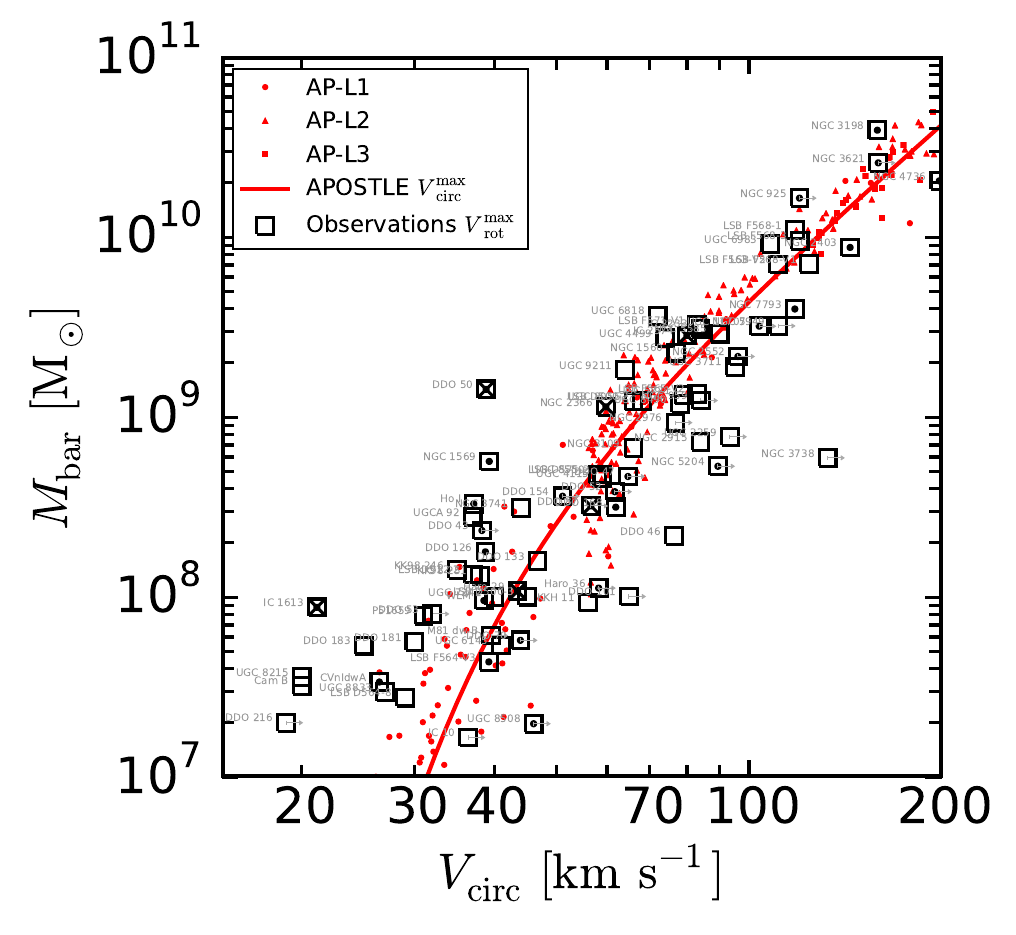}}
  \caption{Baryonic masses of simulated galaxies, $M_{\rm bar}$, as a function of their maximum circular velocity in the APOSTLE simulations (red symbols). Galaxy masses are measured within the galactic radius, defined as $r_{\rm gal}=0.15\, r_{200}$. The thick red solid line shows a fit to the velocity dependence of the median $M_{\rm bar}$ in the simulations. Observed galaxies labelled by their name are shown with open squares and use the maximum measured rotation speed of each galaxy and their baryonic masses, taken from the literature (see \S\ref{SecObsGx} for details on the sample). Squares containing dots correspond to galaxies with rotation curves extending out to at least twice the stellar half-mass radius (see \S\ref{SubsecCores} and Fig.~\ref{FigMbarVhalf}). Squares with crosses highlight the galaxies whose rotation curves are shown in Fig.~\ref{FigRCExamples}.\label{FigMbarVmax}}
\end{figure}

For dark matter-dominated galaxies, the most reliable measure of virial mass is their asymptotic maximum rotation velocity. We therefore begin our analysis by presenting, in Fig.~\ref{FigMbarVmax}, the baryonic mass of APOSTLE galaxies (small red symbols) as a function of the maximum circular velocity, $V_{\rm circ}^{\rm max}$, measured within the virial radius. This figure combines results from the three APOSTLE resolution levels, using only those galaxies whose relevant properties are well resolved (see \S\ref{SecSimGx}).

Baryonic mass correlates strongly and with little scatter with $V_{\rm circ}^{\rm max}$ in simulated galaxies; indeed, the dispersion about the fit\footnote{The functional form of the fit shown in Fig.~\ref{FigMbarVmax} is $M_{\rm bar}/{\rm M}_\odot=5.12 \times 10^9 \, \nu^{3.08} \exp(-0.16\nu^{-2.43})$, where $\nu$ is the maximum circular velocity expressed in units of $100$~km~s$^{-1}$.} shown by the thick solid line is only $0.33$ dex in mass, or $0.08$ dex in velocity. This baryonic Tully-Fisher (BTF) relation is, on average, in remarkable agreement with that of the observed galaxy sample (open black squares), for which we adopt the maximum speed reached by the rotation\footnote{On a technical note, for observed galaxies we actually use the maximum circular velocity estimated from 2D velocity fields as provided by the authors, which typically correct rotation speeds for inclination, asymmetric drift, and instrumental effects. We distinguish these from circular velocities of simulated galaxies, which are estimated directly from the enclosed mass profile, $V_{\rm circ}^2(r)=GM(<r)/r$.} curve of a galaxy, $V_{\rm rot}^{\rm max}$. 

The agreement is encouraging, especially since the APOSTLE simulations use the same code as the EAGLE project, which was calibrated to reproduce the observed number and size of galaxies of stellar mass larger than $\sim 10^{8}\,{\rm M}_\odot$ as a function of stellar mass. Fig.~\ref{FigMbarVmax} thus shows that $\Lambda$CDM simulations that match those constraints also reproduce both the zero-point and velocity scaling of the BTF relation without further calibration. 

One difference, however, seems clear: the scatter in the observed BTF relation appears to increase toward less massive objects, exceeding the rather narrow dispersion about the median trend of the APOSTLE galaxies \citep[see][for a similar conclusion]{PapastergisShankar2015}. We shall discuss the faint end of the simulated BTF relation in a companion paper (Sales et al., in preparation), and focus here on the origin and cosmological significance of the outliers to the BTF relation seen in Fig.~\ref{FigMbarVmax}. Although the existence of such outliers has in the past been regarded with scepticism and ascribed to inferior data, the situation has now changed, and a number of authors have argued that the scatter in the BTF relation genuinely increases toward fainter objects \citep[see, e.g.,][]{Geha2006,Trachternach2009}. The scatter in the {\it inclination-corrected} velocities of observed galaxies shown in Fig.~\ref{FigMbarVmax} increases from $\sim 0.08$ dex to $\sim 0.17$ dex above/below a baryonic mass of $2\times 10^9\, {\rm M}_\odot$. This is much greater than the circular velocity scatter of simulated galaxies, which is $0.04$ dex and $0.05$ dex, respectively, above/below the same baryonic mass.

\subsection{Galaxy formation efficiency}
\label{SubsecFeff}

\begin{figure}
{\leavevmode \includegraphics[width=\columnwidth]{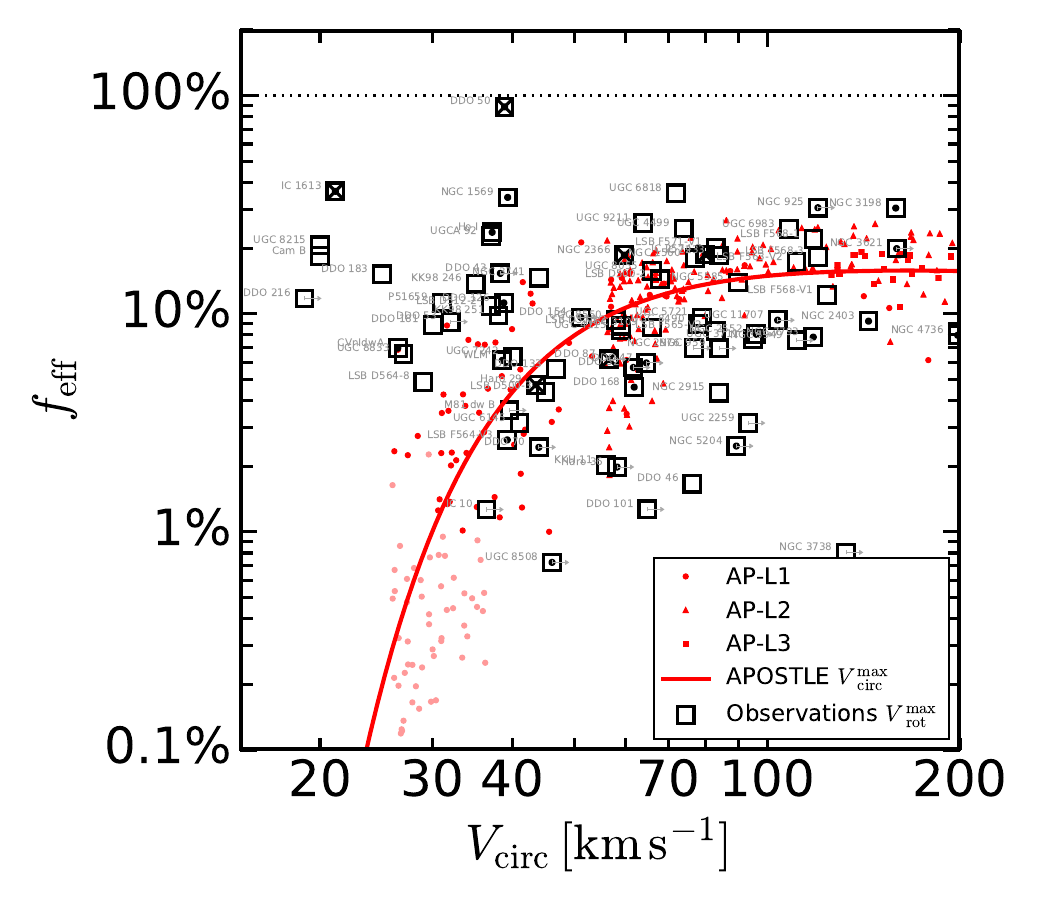}}
\caption{Galaxy formation efficiency, $f_{\rm eff}=M_{\rm bar}/(f_{\rm bar}\, M_{200})$, as a function of maximum circular velocity. Symbols are as in Fig.~\ref{FigMbarVmax}; small red symbols correspond to APOSTLE simulated galaxies (pale symbols have $M_{\rm bar} < 10^7\,{\rm M}_\odot$ and so do not appear in Fig.~\ref{FigMbarVmax}); open squares are observed galaxies. Note that $f_{\rm eff}$ in a simulated galaxy never exceeds $30\%$, but that a number of outliers with anomalously high galaxy formation efficiencies are seen in the observed sample. \label{FigFeffVmax}}
\end{figure}

Examples of BTF outliers -- two of the galaxies highlighted with crosses in Figs.~\ref{FigMbarVmax},~\ref{FigFeffVmax} and \ref{FigMbarVhalf} -- are provided by DDO~50 ($M_{\rm bar}=1.43 \times 10^9\, {\rm M}_\odot$, $V_{\rm rot}^{\rm max}=38.8$~km~s$^{-1}$) and IC~1613 ($M_{\rm bar}=8.77\times 10^8\, {\rm M}_\odot$, $V_{\rm rot}^{\rm max}=19.3$~km~s$^{-1}$), two nearby dwarf galaxies that have been comprehensively studied as part of the LITTLE THINGS survey. These are systems whose baryonic masses are much higher than expected for their velocities or, equivalently, whose measured velocities are much lower than expected for their mass. 

This may be seen in Fig.~\ref{FigFeffVmax}, where we show $f_{\rm eff}$ as a function of $V_{\rm circ}^{\rm max}$ for APOSTLE galaxies compared with observations.  For the latter we plot the maximum observed rotation velocity, and estimate $f_{\rm eff}$ using the best-fitting relation between virial mass and maximum circular velocity derived from the simulations: $M_{200}/{\rm M}_\odot=1.074\times 10^{5} (V_{\rm max}/{\rm km}\;{\rm s}^{-1})^{3.115}$.  As expected from the discussion in \S\ref{SecIntro}, $f_{\rm eff}$ peaks at $\sim 15\%$ for circular velocities comparable to the Milky Way ($\sim 200$~km~s$^{-1}$) but declines precipitously\footnote{The EAGLE hydrodynamics model used in APOSTLE does not include a cold gas phase and therefore does not model molecular hydrogen cooling. This artificially suppresses star formation in small haloes before cosmic reionization, so some of the dwarfs in our simulations have unrealistically low stellar masses -- the decline in $f_{\rm eff}$ may be slightly less abrupt than our results suggest.} toward lower masses, dipping to less than $1\%$ for haloes below $30$~km~s$^{-1}$. If the rotation velocities of DDO~50 and IC~1613 trace reliably the maximum circular velocity of their dark matter haloes then these outliers would correspond to systems where the galaxy formation efficiency, $f_{\rm eff}$, is extraordinarily high, at $89\%$ and $36\%$, respectively, despite their low rotation speeds.

Those two galaxies are not the only outliers from the trend predicted by the numerical simulations. There are also systems that fall well {\it below} the solid red curve in Fig.~\ref{FigFeffVmax} and correspond to systems with unexpectedly high rotation velocities for their mass. There are three broad scenarios that could explain these outliers. They may be systems with unusually low galaxy formation efficiency, perhaps as a result of heating by ionizing background radiation, of particularly effective stellar feedback following a strong past starburst, or of environmental effects such as cosmic web stripping \citep{Benitez-Llambay2013}. They may also be galaxies where the baryonic component is heavily concentrated and dominates the potential in the central regions, raising the local circular velocity above the halo asymptotic value. This scenario does not arise in APOSTLE, since the equation of state chosen for the star-forming gas imposes a minimum size for the stellar component of dwarfs \citep[see, e.g., the discussion in \S4.1.2 of][]{Crain2015}. All APOSTLE dwarfs are dark matter dominated; heavily concentrated, high-surface brightness dwarfs such as, e.g., M~32, are absent from the simulated sample.

Outliers well {\it above} the thick solid line in Fig.~\ref{FigFeffVmax}, like DDO~50 and IC~1613, are more difficult to explain. The increase in scatter in $f_{\rm eff}$ toward lower masses seen in the simulations does not seem to help, since it mainly adds galaxies with small efficiencies. Indeed, we find {\it no} simulated galaxy where the efficiency exceeds $27\%$ over the whole halo mass range spanned by the simulations. DDO~50, on the other hand, is so massive that over $90\%$ of its available baryons must have been able to cool and assemble at the centre of the halo. This corresponds to roughly $25$ times the average efficiency expected for its circular velocity. The discrepancy is even more dramatic for IC~1613, whose estimated efficiency is $\sim 40\%$ -- the simulation average for its velocity is much less than $1\%$.

Galaxies like DDO~50 and IC~1613 are therefore genuinely puzzling systems for which we find no counterparts in the APOSTLE simulations. If $\Lambda$CDM is the correct structure formation model, then such galaxies indicate that either (i) the simulations are at fault, perhaps grossly underestimating the mean efficiency and scatter in low-mass halos, or that (ii) the observed velocities of faint galaxies are not accurate indicators of the mass of their surrounding halos. 

We are not aware of {\it any} $\Lambda$CDM-motivated model of galaxy formation (semi-analytic or numerical) that can accommodate mean efficiencies as high as those shown in Fig.~\ref{FigFeffVmax} for galaxies with maximum rotation speeds in the range $20$-$40$ km/s without dramatically overpredicting the number of dwarfs. If galaxies as massive as $\sim 10^7 \, M_\odot$ could indeed form in $\sim 20$ km/s halos, then we would expect about $200$ at least as massive within $2$ Mpc of the Local Group barycenter \citep[see, e.g., Fig.~4 in][]{Sawala2016b} when, in fact, there are only $\sim 20$ such galaxies in such volume.  It is also clear from Fig.~\ref{FigFeffVmax} that the disagreement would be much easier to explain if velocities rather than efficiencies were systematically affected, since a factor of two shift in velocity implies a change in inferred efficiency of nearly an order of magnitude. We explore this possibility further below.

\section{Rotation curves and halo masses}
\label{SecRotCHM}

\subsection{Rising rotation curves?}
\label{SubsecRising}

\begin{figure*}
  {\leavevmode \includegraphics[width=1.90\columnwidth]{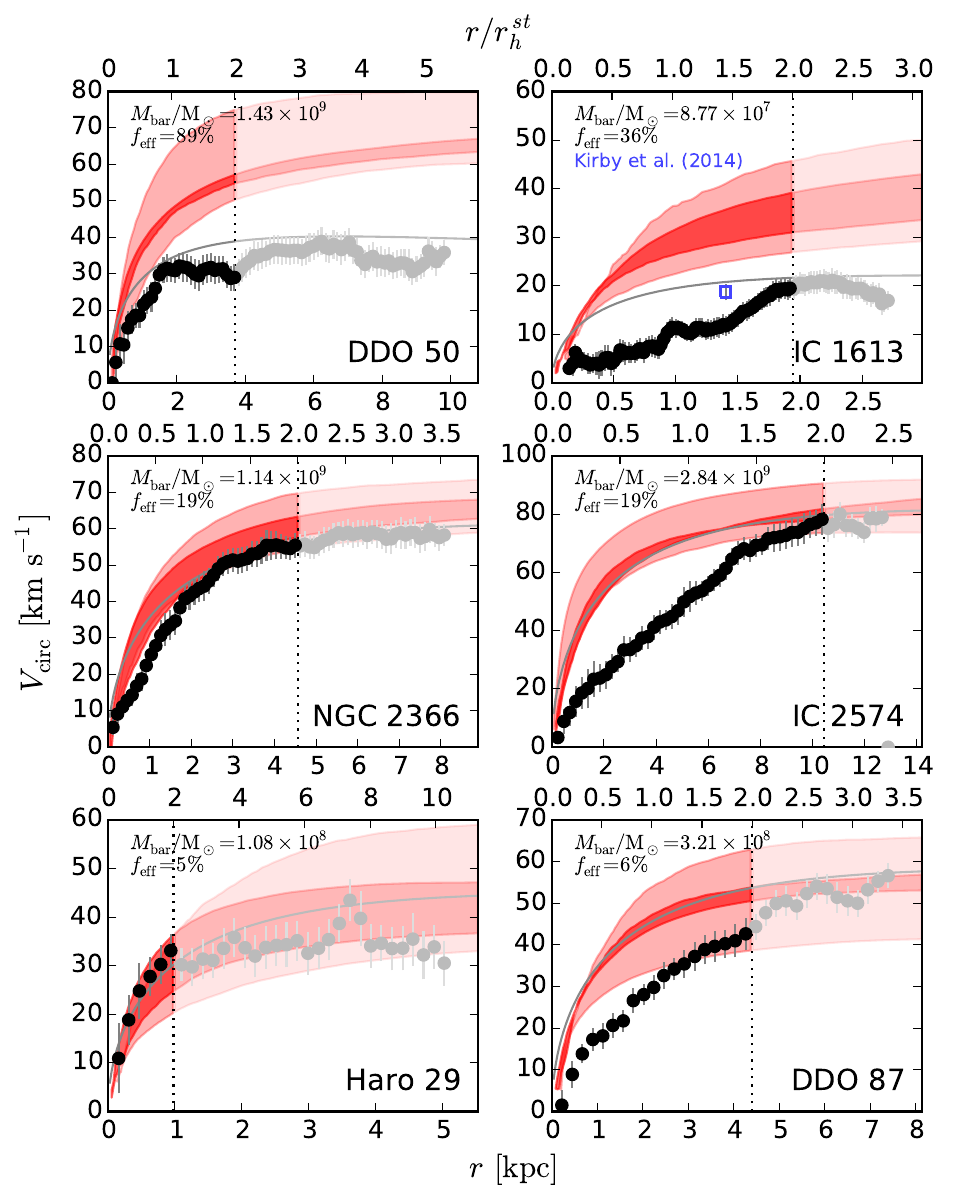}}
  \caption{Examples of galaxies with rotation curves that extend beyond twice the stellar half-mass radius, $r_h^{st}$. These six galaxies are marked with a cross in Figs.~\ref{FigMbarVmax}--\ref{FigMbarVhalf}. In each panel the horizontal axis shows the radius in units of kpc (bottom axis scale) and stellar half-mass radius, $r_h^{st}$ (top axis scale). Thin grey lines show, for reference, the $\Lambda$CDM (NFW) circular velocity profiles of haloes that match the observed maximum rotation speed of each galaxy. The dark and light red-shaded areas indicate the interquartile and full range, respectively, of $V_{\rm circ}$ profiles of the $12$ simulated galaxies whose baryonic masses most closely match that of the galaxy shown in each panel. We highlight the region that contains most of the stars in each galaxy (i.e., $r<2\, r_h^{\rm st}$) with a darker tint. Outside this radius, baryons are not expected to be able to modify the dark matter profile. The top two galaxies are examples of outliers in the velocity-mass relation: these galaxies are anomalously deficient in dark matter (given their baryonic mass). The bottom four galaxies have `normal' galaxy formation efficiency parameters but differ in their inner circular velocity profiles. Those in the left column have rotation curve shapes largely consistent with $\Lambda$CDM haloes of matching maximum velocity. Those on the right show the inner deficit of dark matter at the stellar half-mass radius that is usually associated with a core. For IC~1613 (top right), an independent estimate of the mass in the inner $1.4$~kpc by \citet{Kirby2014} is shown with an open blue symbol.\label{FigRCExamples}}
\end{figure*}

Could the maximum rotation velocity somehow underestimate the asymptotic circular velocity of its surrounding halo? This would be the case, for example, for a galaxy with a rotation curve that is still rising at its last measured point, but it does not apply to either one of the two outliers highlighted above. Indeed, the rotation curves of both DDO~50 and IC~1613 show clear signs of having reached their maximum values (see top panels of Fig.~\ref{FigRCExamples}). That of DDO~50 is a particularly good example, rising quickly to reach its peak and staying flat between $2$ and $10$ kpc.

\subsection{The effects of baryon-induced dark matter `cores'}
\label{SubsecCores}

Another possibility is that baryons might have carved a `core' in the dark matter, thus reducing its central density and, consequently, the circular velocity in the central regions. This creates an inner deficit of dark matter compared with cuspy CDM haloes, which are well approximated by the NFW profile \citep[][]{Navarro1996,Navarro1997}. The characteristic signature of this effect is a rotation curve that rises more gradually near the centre than the sharp rise expected for an NFW profile. 

\begin{figure}
\leavevmode \includegraphics[width=\columnwidth]{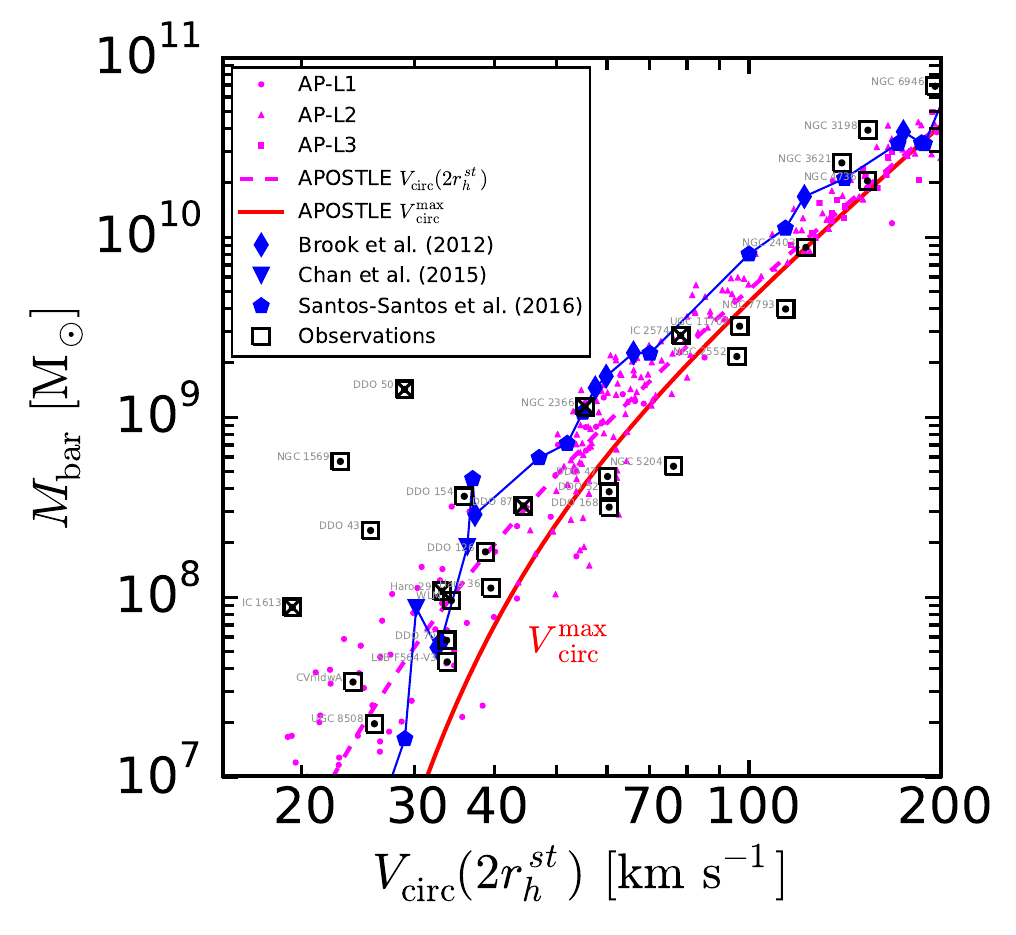}
\caption{As Fig.~\ref{FigMbarVmax}, but for the circular velocity, $V_{\rm circ}(2r_h^{st})$, estimated at twice the stellar half-mass radius. The magenta small filled symbols and thick dashed line correspond to APOSTLE simulated galaxies. The thick red solid line indicates, as in Fig.~\ref{FigMbarVmax}, the results for the maximum circular velocity, and is included for reference only. Open squares correspond to all galaxies in our observed sample where the rotation curve extends at least as far as $2r_h^{st}$. The larger blue solid symbols (connected by a thin line) are individual simulated galaxies where the formation of a core in the central dark matter distribution has been reported. \label{FigMbarVhalf}}
\end{figure}

We examine this possibility in Fig.~\ref{FigMbarVhalf}, where we show again the baryonic Tully-Fisher relation but using, for both simulated and observed galaxies, the circular velocity at the outskirts of the luminous galaxy -- i.e., at twice the stellar half-mass radius, $V_{\rm rot}(2\, r_h^{st})$ -- rather than its maximum attained value.  This choice is useful because
velocities measured as far from the centre as $\sim 2\, r_h^{st}$ should also be largely unaffected by the presence of a possible baryon-induced core. This is because, at least for the core formation mechanism discussed by \citet{Pontzen2014}, the effects of baryons on the dark matter mass profile is largely limited to the regions of a galaxy where stars form.

This is confirmed by the connected symbols in Fig.~\ref{FigMbarVhalf}, which indicate results for $22$ simulated galaxies where a baryon-induced core in the dark matter has been reported in the literature \citep[these have been selected from][]{Brook2012,Santos-Santos2016,Chan2015}. The magenta symbols in the same figure show the results for APOSTLE galaxies, which show no evidence for a core \citep{Schaller2015,Oman2015}. As may be seen from the slight shift between the connected line and the magenta dashed line, cores induce a slight reduction in the circular velocity at $2\, r_h^{st}$, but the changes do not exceed $20\%$ relative to APOSTLE, even for the most extreme examples. Galaxies like DDO~50 or IC~1613 are still extreme outliers that remain unaccounted for, even in simulations with cores.

\subsection{Missing dark matter?}
\label{SecMissDM}

Rather than anomalously baryon rich, galaxies like DDO~50, IC~1613 (the leftmost outliers in Fig.~\ref{FigMbarVhalf}) could alternatively be considered as anomalously low in their dark matter content. These galaxies would then have low circular velocities because they would be `missing dark matter', a result reminiscent of the inner deficit of cold dark matter that characterizes dwarfs where a core has been inferred from their inner rotation curves \citep[for a full discussion, see][]{Oman2015}. From this perspective, galaxies like the four aforementioned outliers would be simply systems where the dark mass deficit is not restricted to the inner regions but rather applies to the whole radial extent of the luminous galaxy, and beyond.

This is illustrated in the top two panels of Fig.~\ref{FigRCExamples}, where we compare the rotation curves of DDO~50 and IC~1613 with the circular velocity profiles of APOSTLE galaxies of matching $M_{\rm bar}$, which are shown bracketed by the red shaded areas. These two systems are clearly missing dark matter from the entire body of the galaxy if their galaxy formation efficiency is comparable to that in simulations. The differences are not subtle. For DDO~50, the comparison implies a total deficit of roughly $\sim 8\times 10^9\, {\rm M}_\odot$ from the inner $10$~kpc, almost an order of magnitude greater than the baryonic mass of the galaxy itself.

The case of DDO~50 and IC~1613 also illustrates that unusually high galaxy formation efficiencies do not occur solely in galaxies with slowly-rising rotation curves, where the presence of a core in the central dark matter distribution might be suspected. This may be seen by considering the thin grey lines in Fig.~\ref{FigRCExamples}, which indicate the expected mass profiles of $\Lambda$CDM haloes \citep[i.e., NFW profiles with average concentration for that cosmology, see, e.g.,][]{Ludlow2014} chosen to match the observed maximum rotation velocity. IC~1613 shows clearly the inner mass deficit ascribed to a core: at $r=r_h^{st}\sim 1$~kpc, the predicted circular velocity exceeds the measured value by nearly a factor of $2$. On the other hand, DDO~50 shows no evidence for a prominent core; its rotation curve rises sharply and flattens out just as expected for a $\Lambda$CDM halo.

The other four galaxies shown in Fig.~\ref{FigRCExamples} provide further examples of the disconnect between inner cores and galaxy formation efficiency. These galaxies have been chosen to span a wide range in $f_{\rm eff}$, decreasing from top to bottom. Those on the right have rotation curves with clear signs of an inner core, whereas those on the left are reasonably well fit by cuspy NFW profiles (thin grey lines) over their full radial extent. The rotation curves of all galaxies in our sample where the rotation curve extends to at least $2r_h^{st}$ are shown in the Appendix.

The anomalies in the galaxy formation efficiency highlighted above thus seem to occur regardless of the inferred presence of a core. In the context of $\Lambda$CDM this implies that a mechanism that allows the galaxy formation efficiency in dwarfs to vary wildly at fixed halo mass is needed in order to understand these observations. It also implies that it is unlikely that these two puzzles can be explained away by a single mechanism, such as baryon-induced cores in the central structure of dark haloes. Resolving these puzzles would thus seem to require the inclusion of some additional physics still missing from simulations of dwarf galaxy formation in $\Lambda$CDM. 

\subsection{Observational and modelling uncertainties}

Before entertaining more far-fetched explanations of the puzzles discussed above, we explore a few more prosaic possibilities. These include the possibility that (i) erroneous galaxy distances have led to substantial overestimation of their baryonic masses (which scale with the assumed distance squared); (ii) that some of the dark matter has been tidally stripped by interaction with a more massive neighbour; and (iii) that the inclination of the galaxies has been overestimated, leading to substantial underestimation of their true rotation speeds.

A thorough analysis of these possible explanations for the full observed sample is beyond the scope of this paper, but we have checked whether such concerns apply to DDO~50 and IC~1613, two clear outliers from the relations discussed above.

\subsubsection{Distances}

The distances to both galaxies seem quite secure: both have distances measured using multiple precise estimators. The apparent luminosity of Cepheids in DDO~50 yields a distance estimate of $3.05\pm0.21$~Mpc \citep{Hoessel1998}, and \emph{Hubble Space Telescope} (\emph{HST}) photometry gives a tip of the red giant branch (TRGB) distance estimate of $3.38\pm0.05$~Mpc \citep{Dalcanton2009}. IC~1613 has similarly high-quality data, with \emph{HST}-based Cepheid and TRGB distance estimates of $0.77\pm0.04$ and $0.71\pm0.06$~Mpc, respectively \citep{Ferrarese2000}. These distances are in good agreement with those assumed by \citet[][$3.4$~Mpc for DDO~50 and $0.7$~Mpc for IC~1613]{Oh2015}. The errors in the distances required to reconcile the baryonic masses of these galaxies with our BTF relation are extreme. For instance, to reduce the inferred mass of DDO~50 by the order of magnitude needed to make it plausibly consistent with our simulation results would imply a distance of only 1.1~Mpc.

\subsubsection{Stripping}
 It also seems improbable that either DDO~50 or IC~1613 have undergone any substantial dark matter stripping due to a tidal interaction with a massive neighbour. According to the catalogue of nearby galaxies compiled by \citet{Tully2009}, the nearest brighter galaxy to DDO~50 is NGC~2403 at a separation of $373$~kpc. IC~1613 is similarly isolated, with no galaxies brighter than itself closer than M~33, at a separation of $449$~kpc.

\subsubsection{Inclination effects}
\label{SecInc}

Of the effects considered in this section, the estimates of the inclinations of DDO~50 and IC~1613 are perhaps the least secure, although the errors necessary to bring the galaxies into agreement with our predicted efficiencies are much larger than the uncertainties quoted in the literature.

It is well known that rotation curve analyses of galaxies with $i\lesssim 40^\circ$ are compromised by the difficulty of deriving robust inclinations solely from the kinematic data \citep[see, e.g.,][]{Begeman1989,deBlok2008}. Even if a minimum inclination is adopted this might still fail to exclude problematic low-inclination galaxies if their kinematic inclinations somehow suggest much larger values \citep[e.g.][]{Read2016}.

The mean inclination of DDO~50 (also known as Holmberg~II) derived in the tilted-ring analysis of \citet{Oh2015} is $49^\circ.7\pm6^\circ.0$, a relatively high value consistent with that inferred by \citet{BureauCarignan2002} from independent, lower-resolution data, and with the $\sim 47^\circ$ inclination estimated from the shape of the galaxy in the \emph{V}-band \citep{Hunter2012}. The true inclination would need to be of order $20^\circ$ for consistency with our simulation results, implying a correction of order $\sim 30^\circ$, much larger than the quoted uncertainty. Such a low inclination (and hence much larger rotation velocities) has been argued for by \citet[][see also \citealt{SanchezSalcedo2014}]{Gentile2012} after re-analysing the data for DDO~50 presented by \citet{Oh2011}. The \citeauthor{Gentile2012} analysis focuses on the low ellipticity of the outer regions of the HI disk, and was motivated by an attempt to reconcile DDO~50 with the predictions of Modified Newtonian Dynamics (MOND).

More recent evaluation of the same data by \citet{Oh2015}, however, appears to confirm the original inclination estimate, although some oddities remain. These are clearly illustrated by the disk-halo decomposition analysis shown in their fig.~A.15. Note, for example, the {\it decreasing} importance of the dark matter with increasing radius, a result that runs counter the established trend for most galaxies. Indeed, at the outermost radius, where dark matter is usually most prominent, the gas contribution accounts almost fully for the observed velocity and the cumulative dark matter contribution is negligible. These unusual properties cast severe doubts on the robustness of the circular velocities derived for DDO~50.

The inclination of IC~1613 is also suspect. In their tilted-ring analysis, \citet{Oh2015} derive a mean kinematic inclination of $48^\circ\pm0^\circ$. This result, together with the small error quoted, are difficult to reconcile with the fact that, when they allow the inclination to be a free parameter those of individual rings scatter widely between $15^\circ$ and $85^\circ$.  Indeed, the rotation curve shown in Fig.~\ref{FigRCExamples} for IC~1613 assumes an inclination of $35^\circ$; this is probably a compromise choice by the authors which, however, is not justified further.

An inclination of $\sim 20^\circ$ would be sufficient to bring IC~1613 within the scatter of our simulated BTF relation by raising its rotation velocity from $\sim 20$ to $\sim 30$~km~s$^{-1}$. It is difficult to assess whether this is plausible. The geometric inclination obtained from the \emph{V}-band shape of IC~1613 is estimated at $37^\circ.9$ \citep{Hunter2012}, close to the final value adopted by \citeauthor{Oh2015}. On the other hand, an independent estimate of the circular velocity at $\sim 1.4$~kpc may be obtained using the velocity dispersion and half-light radius of the galaxy \citep{Kirby2014}: this technique is insensitive to inclination and gives $18.7^{+1.7}_{-1.6}$~km~s$^{-1}$ suggesting that the circular velocity curve of \citeauthor{Oh2015} should indeed be revised upwards by $\sim 50\%$ (see open blue symbol in the top-right panel of Fig.~\ref{FigRCExamples}).


The preceding discussion, albeit inconclusive for DDO~50 and IC~1613, illustrates that inclination error estimates, as well as degeneracies in the algorithms used to map the circular velocity of a galaxy from 2D velocity fields, should be carefully reviewed and critically examined. One final example makes this point quite clear: NGC~3738 is also an extreme BTF outlier\footnote{NGC~3738 is the farthest right outlier in Fig.~\ref{FigMbarVmax}, at $M_{\rm bar}=5.9\times10^{8}\,{\rm M}_\odot$, $V_{\rm rot}^{\rm max}\sim 133$~km~s$^{-1}$. It is not included in Table~\ref{ObsTable} because of the short radial extent of its available rotation curve.}, but on the {\it opposite} side of the relation shown in Fig.~\ref{FigMbarVmax}. This is a case where the rotation speed is twice as high as expected for its baryonic mass and it could even be higher, since its rotation curve appears to still be rising at the outermost measured point. Taken at face value, this would imply an extremely {\it low} galaxy formation efficiency ($f_{\rm eff}\sim 1\%$, see Fig.~\ref{FigFeffVmax}), perhaps signalling unusually efficient feedback or environmental effects. Or an inclination error. NGC~3738 is a nearly face-on galaxy\footnote{On these grounds it could be argued that this galaxy is unsuitable for a tilted-ring analysis.} with a reported mean inclination of $22^\circ.6\pm0^\circ.1$ \citep{Oh2015}. The rotation curve is derived using an inclination fixed at this mean value, but the inclinations preferred by the initial tilted ring fit with inclination as a free parameter vary between $10^\circ$ and $70^\circ$. If the inclination were instead about $20^\circ$ larger than the reported mean, NGC~3738 would lie within the scatter of the results of our simulations.

\section{Summary and Conclusions} \label{SecConc}

We have analysed the baryonic masses and circular velocities of a sample of galaxies with excellent photometric data and high-quality HI observations and compared them with the results of recent $\Lambda$CDM cosmological hydrodynamical simulations from the APOSTLE project. The simulations used the same code developed for the EAGLE project, where the subgrid feedback physics modules have been calibrated to match the galaxy stellar mass function and stellar size distribution of galaxies more massive than the great majority of those studied in this paper.

Our main conclusions may be summarized as follows.

\begin{itemize}

\item The correlation between maximum circular velocity and baryonic mass (the `baryonic Tully-Fisher', or BTF relation) of simulated galaxies reproduces well the zero-point and velocity scaling of observed galaxies in the range $(30,200)$ km s$^{-1}$. This implies that $\Lambda$CDM galaxies of the right size and mass can match naturally the main trends of the BTF relation without further tuning.

\item The sizeable scatter in the observed BTF relation at the faint end, on the other hand, is at odds with the tight relation predicted by our simulations. Particularly challenging are dwarf galaxies, where---taking the data at face value---high baryonic masses and low rotation velocities imply halo masses so low that the inferred efficiency of galaxy formation is extraordinarily high (up to nearly 100\%). We find {\it no} counterparts to such galaxies in APOSTLE.

\item Alternately, these could be systems with anomalously low dark matter content. We demonstrate that this `missing dark matter' cannot be ascribed to the presence of a core, since the mass deficit extends over the whole luminous radius of the affected galaxies, and beyond. Furthermore, `missing dark matter' galaxies include several examples where the rotation curves do not suggest a core, and viceversa.

\item No model of galaxy formation that we are aware of can reconcile these `missing dark matter' systems with $\Lambda$CDM; if such observations hold, they would signal the need for radical modification in our understanding of dwarf galaxy formation in $\Lambda$CDM.

\item Close examination of the data, however, suggest a more plausible explanation, where outliers to our simulated BTF are simply nearly face-on galaxies where the inclinations have been overestimated, and the inclination errors have been substantially underestimated. 

\end{itemize}

If inclination errors are truly responsible for the outliers from the BTF relation, then the {\it outer} dark mass deficits of `missing dark matter' galaxies and the {\it inner} mass deficits (usually ascribed to `cores') explored in \citet{Oman2015} may just be two manifestations of the shortcomings of `tilted-ring' models that attempt to extract the circular velocity profile from gas velocity fields, especially in dwarf irregular galaxies. Continued efforts to understand the limitations of such models, especially using mock observations of realistic simulations of dwarf irregulars, where model output and known input can be compared directly, will be critical to making real progress in confirming or refuting this explanation.

This discussion suggests that caution must be exercised when comparing the mass measurements of dwarf galaxies with simulation results. BTF outliers have featured in discussions of the `too-big-to-fail' problem and of the `core-cusp' issue \citep{BoylanKolchin2012,GarrisonKimmel2014,Papastergis2015,Flores1994,Moore1994,Pontzen2014}. If the cases of DDO~50 and IC~1613 are any guide, their mass profiles might be much more uncertain than the quoted errors would suggest.  

This note of caution applies not only to mass profiles inferred from gas velocity fields, but also to Jeans-estimates of the mass enclosed within the stellar half-mass radius based on stellar velocity dispersions \citep{Walker2009,Wolf2010}. A recent analysis by \citet{Campbell2016} shows that the precision of such estimators is no better than $\sim 20\%$, even when the errors in the half-mass radii and velocity dispersions are significantly smaller. Increased errors would substantially alleviate many of the perceived problems of $\Lambda$CDM on dwarf-galaxy scales.

On the other hand, should future data/analysis confirm the existence of BTF outliers like the ones discussed above, the severity of the `missing dark matter' problem, together with the apparent failure of `baryon physics' to solve it, might motivate the consideration of more radical solutions. One worth highlighting is that the diversity may reflect some intrinsic particle-physics property of the dark matter. This is the case of `self-interacting' dark matter, where, it has been argued, sizeable dispersion in the inner regions of dark matter haloes of given mass may result from scatter in their assembly history \citep[see, e.g.,][and references therein]{Kaplinghat2015}. No detailed simulations of this process are available yet on dwarf galaxy scales, but it is certainly a possibility that needs to be developed further. 

It remains to be seen whether the `missing dark matter' problem points to `missing physics' or `modelling misses'. 
Regardless, we are hopeful that the puzzles outlined above will be profitably used to help guide future developments in our understanding of dwarf galaxy formation.

\section*{Acknowledgements}\label{sec-awknowledgements}

We thank S.-H.~Oh, E.~de~Blok, C.~Brook and J.~Adams for data contributions. This work was supported by the Science and Technology Facilities Council (grant number ST/F001166/1). CSF acknowledges ERC Advanced Grant 267291 COSMIWAY and JFN a Leverhulme Visiting Professor grant held at the Institute for Computational Cosmology, Durham University. This work used the DiRAC Data Centric system at Durham University, operated by the Institute for Computational Cosmology on behalf of the STFC DiRAC HPC Facility (www.dirac.ac.uk). This equipment was funded by BIS National E-infrastructure capital grant ST/K00042X/1, STFC capital grant ST/H008519/1, and STFC DiRAC Operations grant ST/K003267/1 and Durham University. DiRAC is part of the National E-Infrastructure. This research has made use of NASA's Astrophysics Data System. This research has made use of the NASA/IPAC Extragalactic Database (NED) which is operated by the Jet Propulsion Laboratory, California Institute of Technology, under contract with the National Aeronautics and Space Administration.

\bibliography{paper}

\begin{thebibliography}{}
\expandafter\ifx\csname natexlab\endcsname\relax\def\natexlab#1{#1}\fi

\bibitem[{{Adams} {et~al.}(2014){Adams}, {Simon}, {Fabricius}, {van den Bosch},
  {Barentine}, {Bender}, {Gebhardt}, {Hill}, {Murphy}, {Swaters}, {Thomas}, \&
  {van de Ven}}]{Adams2014}
{Adams}, J.~J., {Simon}, J.~D., {Fabricius}, M.~H., {et~al.} 2014, \apj, 789,
  63

\bibitem[{{Begeman}(1989)}]{Begeman1989}
{Begeman}, K.~G. 1989, \aap, 223, 47

\bibitem[{{Behroozi} {et~al.}(2013){Behroozi}, {Wechsler}, \&
  {Conroy}}]{Behroozi2013}
{Behroozi}, P.~S., {Wechsler}, R.~H., \& {Conroy}, C. 2013, \apj, 770, 57

\bibitem[{{Ben{\'{\i}}tez-Llambay} {et~al.}(2013){Ben{\'{\i}}tez-Llambay},
  {Navarro}, {Abadi}, {Gottl{\"o}ber}, {Yepes}, {Hoffman}, \&
  {Steinmetz}}]{Benitez-Llambay2013}
{Ben{\'{\i}}tez-Llambay}, A., {Navarro}, J.~F., {Abadi}, M.~G., {et~al.} 2013,
  \apjl, 763, L41

\bibitem[{{Benson} {et~al.}(2002){Benson}, {Lacey}, {Baugh}, {Cole}, \&
  {Frenk}}]{Benson2002}
{Benson}, A.~J., {Lacey}, C.~G., {Baugh}, C.~M., {Cole}, S., \& {Frenk}, C.~S.
  2002, \mnras, 333, 156

\bibitem[{{Binney} \& {Tremaine}(2008)}]{BT2008}
{Binney}, J., \& {Tremaine}, S. 2008, {Galactic Dynamics: Second Edition}
  (Princeton University Press)

\bibitem[{{Boylan-Kolchin} {et~al.}(2012){Boylan-Kolchin}, {Bullock}, \&
  {Kaplinghat}}]{BoylanKolchin2012}
{Boylan-Kolchin}, M., {Bullock}, J.~S., \& {Kaplinghat}, M. 2012, \mnras, 422,
  1203

\bibitem[{{Brook} {et~al.}(2012){Brook}, {Stinson}, {Gibson}, {Wadsley}, \&
  {Quinn}}]{Brook2012}
{Brook}, C.~B., {Stinson}, G., {Gibson}, B.~K., {Wadsley}, J., \& {Quinn}, T.
  2012, \mnras, 424, 1275

\bibitem[{{Bullock} {et~al.}(2000){Bullock}, {Kravtsov}, \&
  {Weinberg}}]{Bullock2000}
{Bullock}, J.~S., {Kravtsov}, A.~V., \& {Weinberg}, D.~H. 2000, \apj, 539, 517

\bibitem[{{Bureau} \& {Carignan}(2002)}]{BureauCarignan2002}
{Bureau}, M., \& {Carignan}, C. 2002, \aj, 123, 1316

\bibitem[{{Campbell} {et~al.}(2016){Campbell}, {Frenk}, {Jenkins}, {Eke},
  {Navarro}, {Sawala}, {Schaller}, {Fattahi}, {Oman}, \&
  {Theuns}}]{Campbell2016}
{Campbell}, D.~J.~R., {Frenk}, C.~S., {Jenkins}, A., {et~al.} 2016, ArXiv
  e-prints, arXiv:1603.04443

\bibitem[{{Chan} {et~al.}(2015){Chan}, {Kere{\v s}}, {O{\~n}orbe}, {Hopkins},
  {Muratov}, {Faucher-Gigu{\`e}re}, \& {Quataert}}]{Chan2015}
{Chan}, T.~K., {Kere{\v s}}, D., {O{\~n}orbe}, J., {et~al.} 2015, \mnras, 454,
  2981

\bibitem[{{Crain} {et~al.}(2015){Crain}, {Schaye}, {Bower}, {Furlong},
  {Schaller}, {Theuns}, {Dalla Vecchia}, {Frenk}, {McCarthy}, {Helly},
  {Jenkins}, {Rosas-Guevara}, {White}, \& {Trayford}}]{Crain2015}
{Crain}, R.~A., {Schaye}, J., {Bower}, R.~G., {et~al.} 2015, \mnras, 450, 1937

\bibitem[{{Dalcanton} {et~al.}(2009){Dalcanton}, {Williams}, {Seth}, {Dolphin},
  {Holtzman}, {Rosema}, {Skillman}, {Cole}, {Girardi}, {Gogarten},
  {Karachentsev}, {Olsen}, {Weisz}, {Christensen}, {Freeman}, {Gilbert},
  {Gallart}, {Harris}, {Hodge}, {de Jong}, {Karachentseva}, {Mateo}, {Stetson},
  {Tavarez}, {Zaritsky}, {Governato}, \& {Quinn}}]{Dalcanton2009}
{Dalcanton}, J.~J., {Williams}, B.~F., {Seth}, A.~C., {et~al.} 2009, \apjs,
  183, 67

\bibitem[{{Dalla Vecchia} \& {Schaye}(2012)}]{DallaVecchia2012}
{Dalla Vecchia}, C., \& {Schaye}, J. 2012, \mnras, 426, 140

\bibitem[{{Davis} {et~al.}(1985){Davis}, {Efstathiou}, {Frenk}, \&
  {White}}]{Davis1985}
{Davis}, M., {Efstathiou}, G., {Frenk}, C.~S., \& {White}, S.~D.~M. 1985, \apj,
  292, 371

\bibitem[{{de Blok} {et~al.}(2008){de Blok}, {Walter}, {Brinks},
  {Trachternach}, {Oh}, \& {Kennicutt}}]{deBlok2008}
{de Blok}, W.~J.~G., {Walter}, F., {Brinks}, E., {et~al.} 2008, \aj, 136, 2648

\bibitem[{{Dolag} {et~al.}(2009){Dolag}, {Borgani}, {Murante}, \&
  {Springel}}]{Dolag2009}
{Dolag}, K., {Borgani}, S., {Murante}, G., \& {Springel}, V. 2009, \mnras, 399,
  497

\bibitem[{{Efstathiou}(1992)}]{Efstathiou1992}
{Efstathiou}, G. 1992, \mnras, 256, 43P

\bibitem[{{Fattahi} {et~al.}(2015){Fattahi}, {Navarro}, {Sawala}, {Frenk},
  {Oman}, {Crain}, {Furlong}, {Schaller}, {Schaye}, {Theuns}, \&
  {Jenkins}}]{Fattahi2015}
{Fattahi}, A., {Navarro}, J.~F., {Sawala}, T., {et~al.} 2015, ArXiv e-prints,
  arXiv:1507.03643

\bibitem[{{Ferrarese} {et~al.}(2000){Ferrarese}, {Ford}, {Huchra}, {Kennicutt},
  {Mould}, {Sakai}, {Freedman}, {Stetson}, {Madore}, {Gibson}, {Graham},
  {Hughes}, {Illingworth}, {Kelson}, {Macri}, {Sebo}, \&
  {Silbermann}}]{Ferrarese2000}
{Ferrarese}, L., {Ford}, H.~C., {Huchra}, J., {et~al.} 2000, \apjs, 128, 431

\bibitem[{{Ferrero} {et~al.}(2012){Ferrero}, {Abadi}, {Navarro}, {Sales}, \&
  {Gurovich}}]{Ferrero2012}
{Ferrero}, I., {Abadi}, M.~G., {Navarro}, J.~F., {Sales}, L.~V., \& {Gurovich},
  S. 2012, \mnras, 425, 2817

\bibitem[{{Flores} \& {Primack}(1994)}]{Flores1994}
{Flores}, R.~A., \& {Primack}, J.~R. 1994, \apjl, 427, L1

\bibitem[{{Garrison-Kimmel} {et~al.}(2014){Garrison-Kimmel}, {Boylan-Kolchin},
  {Bullock}, \& {Lee}}]{GarrisonKimmel2014}
{Garrison-Kimmel}, S., {Boylan-Kolchin}, M., {Bullock}, J.~S., \& {Lee}, K.
  2014, \mnras, 438, 2578

\bibitem[{{Geha} {et~al.}(2006){Geha}, {Blanton}, {Masjedi}, \&
  {West}}]{Geha2006}
{Geha}, M., {Blanton}, M.~R., {Masjedi}, M., \& {West}, A.~A. 2006, \apj, 653,
  240

\bibitem[{{Gentile} {et~al.}(2012){Gentile}, {Angus}, {Famaey}, {Oh}, \& {de
  Blok}}]{Gentile2012}
{Gentile}, G., {Angus}, G.~W., {Famaey}, B., {Oh}, S.-H., \& {de Blok},
  W.~J.~G. 2012, \aap, 543, A47

\bibitem[{{Gottloeber} {et~al.}(2010){Gottloeber}, {Hoffman}, \&
  {Yepes}}]{Gottloeber2010}
{Gottloeber}, S., {Hoffman}, Y., \& {Yepes}, G. 2010, ArXiv e-prints,
  arXiv:1005.2687

\bibitem[{{Haardt} \& {Madau}(2001)}]{Haardt2001}
{Haardt}, F., \& {Madau}, P. 2001, in Clusters of Galaxies and the High
  Redshift Universe Observed in X-rays, ed. D.~M. {Neumann} \& J.~T.~V. {Tran},
  64

\bibitem[{{Hoessel} {et~al.}(1998){Hoessel}, {Saha}, \&
  {Danielson}}]{Hoessel1998}
{Hoessel}, J.~G., {Saha}, A., \& {Danielson}, G.~E. 1998, \aj, 115, 573

\bibitem[{{Hopkins}(2013)}]{Hopkins2013}
{Hopkins}, P.~F. 2013, \mnras, 428, 2840

\bibitem[{{Hu} \& {Sugiyama}(1995)}]{Hu1995}
{Hu}, W., \& {Sugiyama}, N. 1995, \apj, 444, 489

\bibitem[{{Hunter} \& {Elmegreen}(2006)}]{Hunter2006}
{Hunter}, D.~A., \& {Elmegreen}, B.~G. 2006, \apjs, 162, 49

\bibitem[{{Hunter} {et~al.}(2012){Hunter}, {Ficut-Vicas}, {Ashley}, {Brinks},
  {Cigan}, {Elmegreen}, {Heesen}, {Herrmann}, {Johnson}, {Oh}, {Rupen},
  {Schruba}, {Simpson}, {Walter}, {Westpfahl}, {Young}, \&
  {Zhang}}]{Hunter2012}
{Hunter}, D.~A., {Ficut-Vicas}, D., {Ashley}, T., {et~al.} 2012, \aj, 144, 134

\bibitem[{{Kamphuis} {et~al.}(2015){Kamphuis}, {J{\'o}zsa}, {Oh}, {Spekkens},
  {Urbancic}, {Serra}, {Koribalski}, \& {Dettmar}}]{Kamphuis2015}
{Kamphuis}, P., {J{\'o}zsa}, G.~I.~G., {Oh}, S.-.~H., {et~al.} 2015, \mnras,
  452, 3139

\bibitem[{{Kaplinghat} {et~al.}(2015){Kaplinghat}, {Tulin}, \&
  {Yu}}]{Kaplinghat2015}
{Kaplinghat}, M., {Tulin}, S., \& {Yu}, H.-B. 2015, ArXiv e-prints,
  arXiv:1508.03339

\bibitem[{{Kirby} {et~al.}(2014){Kirby}, {Bullock}, {Boylan-Kolchin},
  {Kaplinghat}, \& {Cohen}}]{Kirby2014}
{Kirby}, E.~N., {Bullock}, J.~S., {Boylan-Kolchin}, M., {Kaplinghat}, M., \&
  {Cohen}, J.~G. 2014, \mnras, 439, 1015

\bibitem[{{Komatsu} {et~al.}(2011){Komatsu}, {Smith}, {Dunkley}, {Bennett},
  {Gold}, {Hinshaw}, {Jarosik}, {Larson}, {Nolta}, {Page}, {Spergel},
  {Halpern}, {Hill}, {Kogut}, {Limon}, {Meyer}, {Odegard}, {Tucker}, {Weiland},
  {Wollack}, \& {Wright}}]{WMAP7}
{Komatsu}, E., {Smith}, K.~M., {Dunkley}, J., {et~al.} 2011, \apjs, 192, 18

\bibitem[{{Larson}(1974)}]{Larson1974}
{Larson}, R.~B. 1974, \mnras, 169, 229

\bibitem[{{Ludlow} {et~al.}(2014){Ludlow}, {Navarro}, {Angulo},
  {Boylan-Kolchin}, {Springel}, {Frenk}, \& {White}}]{Ludlow2014}
{Ludlow}, A.~D., {Navarro}, J.~F., {Angulo}, R.~E., {et~al.} 2014, \mnras, 441,
  378

\bibitem[{{Madau} \& {Dickinson}(2014)}]{Madau2014}
{Madau}, P., \& {Dickinson}, M. 2014, \araa, 52, 415

\bibitem[{{Mashchenko} {et~al.}(2006){Mashchenko}, {Couchman}, \&
  {Wadsley}}]{MashchenkoCouchmanWadsley2006}
{Mashchenko}, S., {Couchman}, H.~M.~P., \& {Wadsley}, J. 2006, \nat, 442, 539

\bibitem[{{McGaugh}(2012)}]{McGaugh2012}
{McGaugh}, S.~S. 2012, \aj, 143, 40

\bibitem[{{Moore}(1994)}]{Moore1994}
{Moore}, B. 1994, \nat, 370, 629

\bibitem[{{Navarro} {et~al.}(1996{\natexlab{a}}){Navarro}, {Eke}, \&
  {Frenk}}]{NavarroEkeFrenk1996}
{Navarro}, J.~F., {Eke}, V.~R., \& {Frenk}, C.~S. 1996{\natexlab{a}}, \mnras,
  283, L72

\bibitem[{{Navarro} {et~al.}(1996{\natexlab{b}}){Navarro}, {Frenk}, \&
  {White}}]{Navarro1996}
{Navarro}, J.~F., {Frenk}, C.~S., \& {White}, S.~D.~M. 1996{\natexlab{b}},
  \apj, 462, 563

\bibitem[{{Navarro} {et~al.}(1997){Navarro}, {Frenk}, \& {White}}]{Navarro1997}
---. 1997, \apj, 490, 493

\bibitem[{{Oh} {et~al.}(2011){Oh}, {de Blok}, {Brinks}, {Walter}, \&
  {Kennicutt}}]{Oh2011}
{Oh}, S.-H., {de Blok}, W.~J.~G., {Brinks}, E., {Walter}, F., \& {Kennicutt},
  Jr., R.~C. 2011, \aj, 141, 193

\bibitem[{{Oh} {et~al.}(2015){Oh}, {Hunter}, {Brinks}, {Elmegreen}, {Schruba},
  {Walter}, {Rupen}, {Young}, {Simpson}, {Johnson}, {Herrmann}, {Ficut-Vicas},
  {Cigan}, {Heesen}, {Ashley}, \& {Zhang}}]{Oh2015}
{Oh}, S.-H., {Hunter}, D.~A., {Brinks}, E., {et~al.} 2015, \aj, 149, 180

\bibitem[{{Oman} {et~al.}(2015){Oman}, {Navarro}, {Fattahi}, {Frenk}, {Sawala},
  {White}, {Bower}, {Crain}, {Furlong}, {Schaller}, {Schaye}, \&
  {Theuns}}]{Oman2015}
{Oman}, K.~A., {Navarro}, J.~F., {Fattahi}, A., {et~al.} 2015, \mnras, 452,
  3650

\bibitem[{{Papastergis} {et~al.}(2015){Papastergis}, {Giovanelli}, {Haynes}, \&
  {Shankar}}]{Papastergis2015}
{Papastergis}, E., {Giovanelli}, R., {Haynes}, M.~P., \& {Shankar}, F. 2015,
  \aap, 574, A113

\bibitem[{{Papastergis} \& {Shankar}(2015)}]{PapastergisShankar2015}
{Papastergis}, E., \& {Shankar}, F. 2015, ArXiv e-prints, arXiv:1511.08741

\bibitem[{{Paturel} {et~al.}(2003){Paturel}, {Petit}, {Prugniel}, {Theureau},
  {Rousseau}, {Brouty}, {Dubois}, \& {Cambr{\'e}sy}}]{Paturel2003}
{Paturel}, G., {Petit}, C., {Prugniel}, P., {et~al.} 2003, \aap, 412, 45

\bibitem[{{Planck Collaboration} {et~al.}(2015){Planck Collaboration}, {Ade},
  {Aghanim}, {Arnaud}, {Ashdown}, {Aumont}, {Baccigalupi}, {Banday},
  {Barreiro}, {Bartlett}, \& et~al.}]{Planck2015}
{Planck Collaboration}, {Ade}, P.~A.~R., {Aghanim}, N., {et~al.} 2015, ArXiv
  e-prints, arXiv:1502.01589

\bibitem[{{Pontzen} \& {Governato}(2014)}]{Pontzen2014}
{Pontzen}, A., \& {Governato}, F. 2014, \nat, 506, 171

\bibitem[{{Power} {et~al.}(2003){Power}, {Navarro}, {Jenkins}, {Frenk},
  {White}, {Springel}, {Stadel}, \& {Quinn}}]{Power2003}
{Power}, C., {Navarro}, J.~F., {Jenkins}, A., {et~al.} 2003, \mnras, 338, 14

\bibitem[{{Read} {et~al.}(2016){Read}, {Iorio}, {Agertz}, \&
  {Fraternali}}]{Read2016}
{Read}, J.~I., {Iorio}, G., {Agertz}, O., \& {Fraternali}, F. 2016, ArXiv
  e-prints, arXiv:1601.05821

\bibitem[{{Rix} \& {Bovy}(2013)}]{Rix2013}
{Rix}, H.-W., \& {Bovy}, J. 2013, \aapr, 21, 61

\bibitem[{{Rogstad} {et~al.}(1974){Rogstad}, {Lockhart}, \&
  {Wright}}]{Rogstad1974}
{Rogstad}, D.~H., {Lockhart}, I.~A., \& {Wright}, M.~C.~H. 1974, \apj, 193, 309

\bibitem[{{S{\'a}nchez-Salcedo} {et~al.}(2014){S{\'a}nchez-Salcedo},
  {Hidalgo-G{\'a}mez}, \& {Mart{\'{\i}}nez-Garc{\'{\i}}a}}]{SanchezSalcedo2014}
{S{\'a}nchez-Salcedo}, F.~J., {Hidalgo-G{\'a}mez}, A.~M., \&
  {Mart{\'{\i}}nez-Garc{\'{\i}}a}, E.~E. 2014, \rmxaa, 50, 225

\bibitem[{{Santos-Santos} {et~al.}(2016){Santos-Santos}, {Brook}, {Stinson},
  {Di Cintio}, {Wadsley}, {Dom{\'{\i}}nguez-Tenreiro}, {Gottl{\"o}ber}, \&
  {Yepes}}]{Santos-Santos2016}
{Santos-Santos}, I.~M., {Brook}, C.~B., {Stinson}, G., {et~al.} 2016, \mnras,
  455, 476

\bibitem[{{Sawala} {et~al.}(2014){Sawala}, {Frenk}, {Fattahi}, {Navarro},
  {Theuns}, {Bower}, {Crain}, {Furlong}, {Jenkins}, {Schaller}, \&
  {Schaye}}]{Sawala2016a}
{Sawala}, T., {Frenk}, C.~S., {Fattahi}, A., {et~al.} 2014, ArXiv e-prints,
  arXiv:1406.6362

\bibitem[{{Sawala} {et~al.}(2015){Sawala}, {Frenk}, {Fattahi}, {Navarro},
  {Bower}, {Crain}, {Dalla Vecchia}, {Furlong}, {Helly}, {Jenkins}, {Oman},
  {Schaller}, {Schaye}, {Theuns}, {Trayford}, \& {White}}]{Sawala2016b}
---. 2015, ArXiv e-prints, arXiv:1511.01098

\bibitem[{{Schaller} {et~al.}(2015){Schaller}, {Frenk}, {Bower}, {Theuns},
  {Jenkins}, {Schaye}, {Crain}, {Furlong}, {Dalla Vecchia}, \&
  {McCarthy}}]{Schaller2015}
{Schaller}, M., {Frenk}, C.~S., {Bower}, R.~G., {et~al.} 2015, \mnras, 451,
  1247

\bibitem[{{Schaye}(2004)}]{Schaye2004}
{Schaye}, J. 2004, \apj, 609, 667

\bibitem[{{Schaye} \& {Dalla Vecchia}(2008)}]{Schaye2008}
{Schaye}, J., \& {Dalla Vecchia}, C. 2008, \mnras, 383, 1210

\bibitem[{{Schaye} {et~al.}(2015){Schaye}, {Crain}, {Bower}, {Furlong},
  {Schaller}, {Theuns}, {Dalla Vecchia}, {Frenk}, {McCarthy}, {Helly},
  {Jenkins}, {Rosas-Guevara}, {White}, {Baes}, {Booth}, {Camps}, {Navarro},
  {Qu}, {Rahmati}, {Sawala}, {Thomas}, \& {Trayford}}]{Schaye2015}
{Schaye}, J., {Crain}, R.~A., {Bower}, R.~G., {et~al.} 2015, \mnras, 446, 521

\bibitem[{{Springel}(2005)}]{Springel2005}
{Springel}, V. 2005, \mnras, 364, 1105

\bibitem[{{Springel} {et~al.}(2001){Springel}, {White}, {Tormen}, \&
  {Kauffmann}}]{Springel2001}
{Springel}, V., {White}, S.~D.~M., {Tormen}, G., \& {Kauffmann}, G. 2001,
  \mnras, 328, 726

\bibitem[{{Steigman}(2007)}]{Steigman2007}
{Steigman}, G. 2007, Annual Review of Nuclear and Particle Science, 57, 463

\bibitem[{{Trachternach} {et~al.}(2009){Trachternach}, {de Blok}, {McGaugh},
  {van der Hulst}, \& {Dettmar}}]{Trachternach2009}
{Trachternach}, C., {de Blok}, W.~J.~G., {McGaugh}, S.~S., {van der Hulst},
  J.~M., \& {Dettmar}, R.-J. 2009, \aap, 505, 577

\bibitem[{{Tully} {et~al.}(2009){Tully}, {Rizzi}, {Shaya}, {Courtois},
  {Makarov}, \& {Jacobs}}]{Tully2009}
{Tully}, R.~B., {Rizzi}, L., {Shaya}, E.~J., {et~al.} 2009, \aj, 138, 323

\bibitem[{{Walker} {et~al.}(2009){Walker}, {Mateo}, {Olszewski},
  {Pe{\~n}arrubia}, {Wyn Evans}, \& {Gilmore}}]{Walker2009}
{Walker}, M.~G., {Mateo}, M., {Olszewski}, E.~W., {et~al.} 2009, \apj, 704,
  1274

\bibitem[{{Walter} {et~al.}(2008){Walter}, {Brinks}, {de Blok}, {Bigiel},
  {Kennicutt}, {Thornley}, \& {Leroy}}]{Walter2008}
{Walter}, F., {Brinks}, E., {de Blok}, W.~J.~G., {et~al.} 2008, \aj, 136, 2563

\bibitem[{{Wang} {et~al.}(2015){Wang}, {Han}, {Cooper}, {Cole}, {Frenk}, \&
  {Lowing}}]{Wang2015}
{Wang}, W., {Han}, J., {Cooper}, A.~P., {et~al.} 2015, \mnras, 453, 377

\bibitem[{{White} \& {Frenk}(1991)}]{White1991}
{White}, S.~D.~M., \& {Frenk}, C.~S. 1991, \apj, 379, 52

\bibitem[{{White} {et~al.}(1993){White}, {Navarro}, {Evrard}, \&
  {Frenk}}]{White1993}
{White}, S.~D.~M., {Navarro}, J.~F., {Evrard}, A.~E., \& {Frenk}, C.~S. 1993,
  \nat, 366, 429

\bibitem[{{White} \& {Rees}(1978)}]{White1978}
{White}, S.~D.~M., \& {Rees}, M.~J. 1978, \mnras, 183, 341

\bibitem[{{Wiersma} {et~al.}(2009{\natexlab{a}}){Wiersma}, {Schaye}, \&
  {Smith}}]{Wiersma2009a}
{Wiersma}, R.~P.~C., {Schaye}, J., \& {Smith}, B.~D. 2009{\natexlab{a}},
  \mnras, 393, 99

\bibitem[{{Wiersma} {et~al.}(2009{\natexlab{b}}){Wiersma}, {Schaye}, {Theuns},
  {Dalla Vecchia}, \& {Tornatore}}]{Wiersma2009b}
{Wiersma}, R.~P.~C., {Schaye}, J., {Theuns}, T., {Dalla Vecchia}, C., \&
  {Tornatore}, L. 2009{\natexlab{b}}, \mnras, 399, 574

\bibitem[{{Wolf} {et~al.}(2010){Wolf}, {Martinez}, {Bullock}, {Kaplinghat},
  {Geha}, {Mu{\~n}oz}, {Simon}, \& {Avedo}}]{Wolf2010}
{Wolf}, J., {Martinez}, G.~D., {Bullock}, J.~S., {et~al.} 2010, \mnras, 406,
  1220

\end{thebibliography}

\appendix
\section{Additional rotation curve examples}
\label{SecApp}

In Fig.~\ref{FigAppendix} we show the rotation curves of all observed galaxies whose rotation curves extend to at least $2r_h^{st}$, i.e. the same galaxies as appear in Fig.~\ref{FigMbarVhalf} and Table~\ref{ObsTable}. This serves to illustrate the striking diversity in rotation curve shapes, in addition to the scatter in $V_{\rm circ}(2r_h^{st})$, relative to the results from simulations. We note rotation curves in reasonable agreement with our simulations at all radii (e.g. Haro 29, WLM, DDO 154, NGC 2366, NGC 2403), rotation curves which agree with our simulated rotation curves at $2r_h^{st}$ but have very different shapes (e.g. CVnIdwA, UGC 8508, DDO 126, IC 2574, DDO 87, NGC 4736), rotation curves with shapes similar to those in our simulations but with systematically high (NGC 5204) or low (NGC 1569, DDO 50) velocities at all radii, and rotation curves that have neither shapes nor velocities at $2r_h^{st}$ consistent with our simulations (e.g. IC 1613, UGC 11707, NGC 7793).

\begin{figure*}
  {\leavevmode \includegraphics[width=2.\columnwidth]{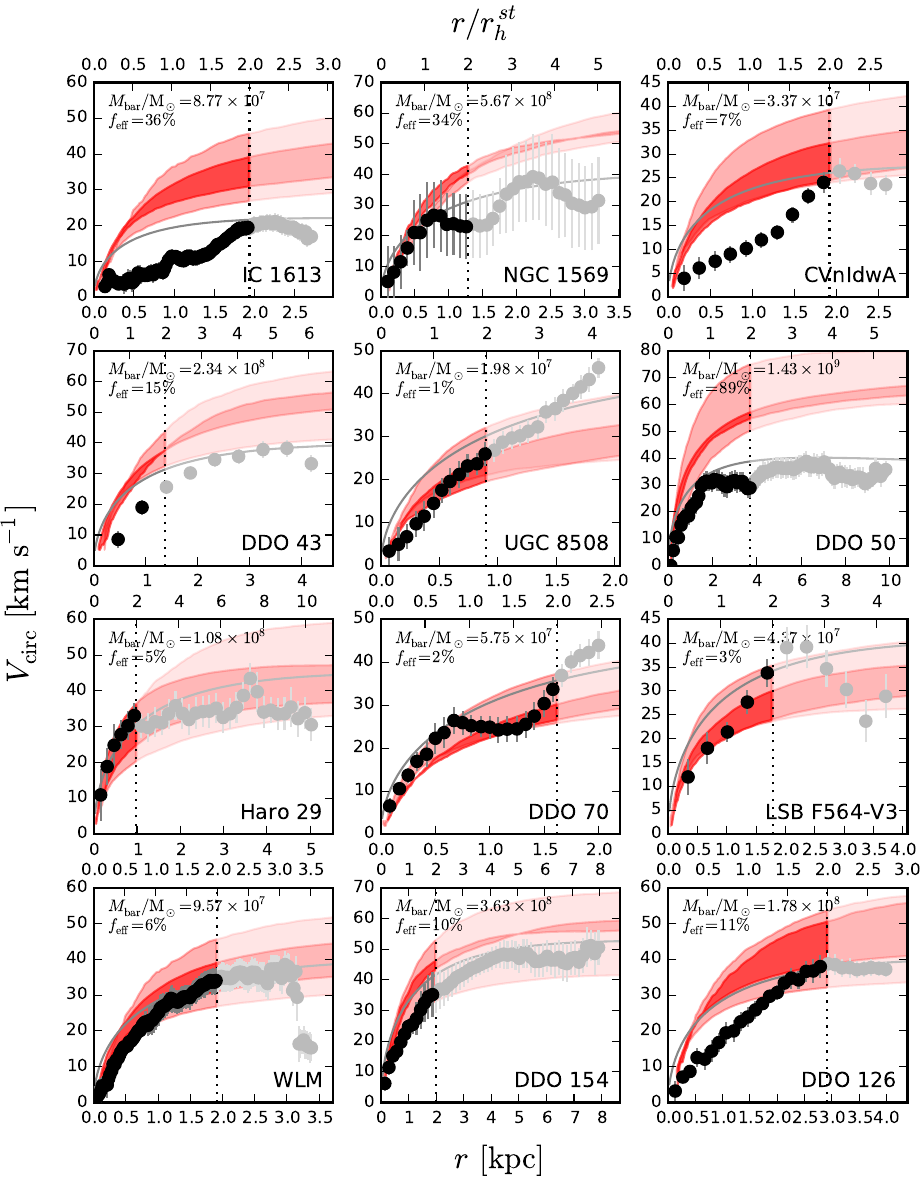}}
  \caption{Rotation curves for all galaxies with rotation curves that extend to at least $2r_h^{st}$ (see also Table~\ref{ObsTable}). The panels are in order of increasing $V_{\rm circ}(2r_h^{st})$. Symbols, lines and shading are as in Fig.~\ref{FigRCExamples}.\label{FigAppendix}}
\end{figure*}

\begin{figure*}
  {\leavevmode \includegraphics[width=2.\columnwidth]{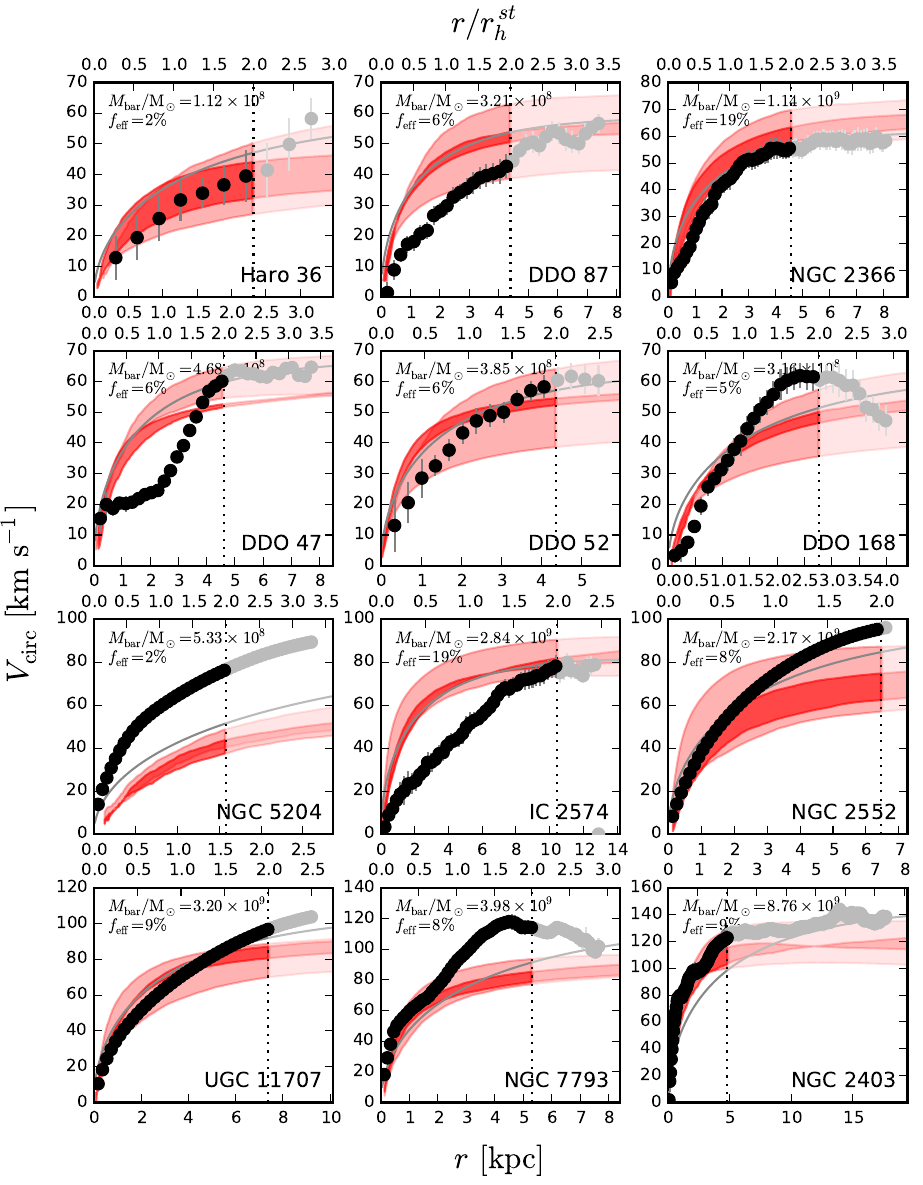}}
  \contcaption{}
\end{figure*}

\begin{figure*}
  {\leavevmode \includegraphics[width=2.\columnwidth]{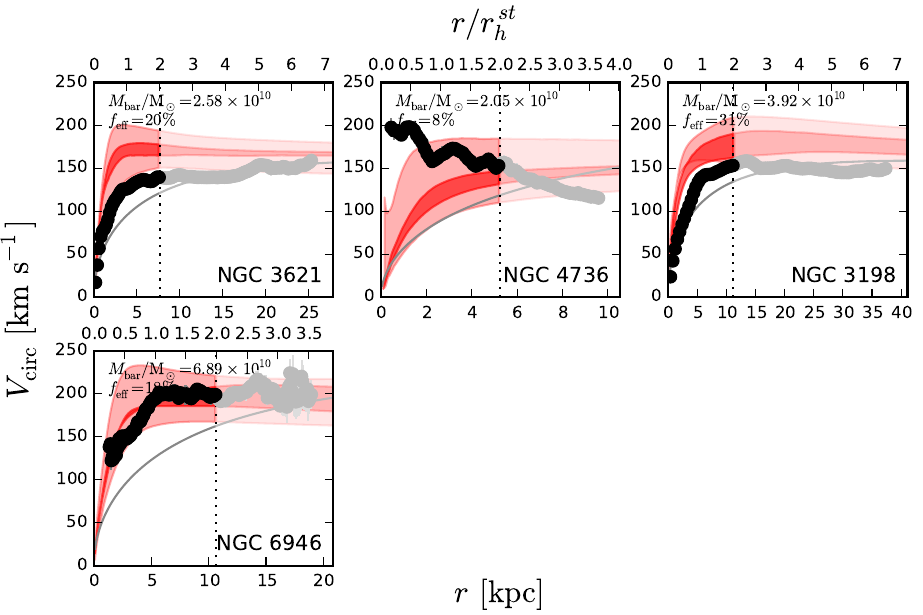}}
  \contcaption{}
\end{figure*}

\label{lastpage}

\end{document}